\documentclass[journal]{IEEEtran}
\usepackage{amsmath,amsfonts}
\usepackage{algorithmic}
\usepackage{algorithm}
\usepackage{array}
\usepackage[caption=false,font=normalsize,labelfont=sf,textfont=sf]{subfig}
\usepackage{textcomp}
\usepackage{stfloats}
\usepackage{url}
\usepackage{verbatim}
\usepackage{graphicx}
\graphicspath{{./fig/}}
\usepackage{enumitem}
\usepackage{xcolor}
\usepackage{breqn}

\usepackage{cite}
\begin{document}

\title{Response Analysis of Four-Level Heterodyne Rydberg Atom Receiver}

\author{Yu Tang, Siyuan Wang, Shuang Ren, Chuang Yang, Hanbin Zhou, and Chenxi Lu
\thanks{Yu Tang, Siyuan Wang, Shuang Ren, Hanbin Zhou, Chuang Yang, and Chenxi Lu are with the National Key Laboratory of Radar Signal Processing and the Institute of Information Sensing, Xidian University, Xi’an 710071, China (e-mail: ytang@xidian.edu.cn).}
\thanks{This work was supported by the National Key Laboratory of Radar Signal Processing under Grant KGJ202101.}
\thanks{Yu Tang is the corresponding author.}}

\markboth{IEEE Transactions on Antennas and Propagation}
{Yu Tang, Siyuan Wang, \MakeLowercase{\textit{et al.}}: Response Analysis of Four-Level Heterodyne Rydberg Atom Receiver}


\maketitle

\begin{abstract}
The four-level heterodyne Rydberg atom receiver has garnered significant attention in microwave detection and communication due to its high sensitivity and phase measurement capabilities. Existing theoretical studies, primarily based on static solutions, are limited in characterizing the system's frequency response. To address this, this paper comprehensively investigates the dynamic solutions of the density matrix elements for the four-level heterodyne structure, establishing a quantitative relationship between system response, signal frequency, and system parameters. This enables theoretical bandwidth calculations and performance analysis. This paper also constructs a noise model for the density matrix elements, revealing the relationship between the ultimate sensitivity of the Rydberg atom receiver and the noise in the density matrix elements. Both theoretical simulation and experimental results demonstrate that the bandwidth of the four-level heterodyne receiver can exceed 10 MHz. This study provides critical theoretical support for the engineering applications and performance optimization of heterodyne Rydberg atom receivers.
\end{abstract}


\section{Introduction}
\IEEEPARstart{R}{ydberg} atoms, characterized by a high principal quantum number and large dipole moments, exhibit extreme sensitivity to external electromagnetic fields\cite{ref1}. Recent studies have demonstrated their significant potential in high-sensitivity microwave sensing, radar, and terahertz communication\cite{ref2, ref3, ref4, ref5, ref6, ref7}. Atomic sensors based on electromagnetically induced transparency (EIT) and Autler-Townes (A-T) splitting have shown unique advantages in weak-field detection and radio-frequency signal reception\cite{ref8, ref9, ref10, ref11, ref12, ref13, ref14, ref15, ref16, ref17, ref18}. However, their sensitivity and phase measurement capabilities remain limited. The heterodyne Rydberg atom receiver, by introducing a microwave local oscillator signal, enables the interaction of the target microwave signal with the local oscillator signal within the atomic system, generating a beat-frequency signal containing information about the target signal\cite{ref3,ref19}. Compared to receivers based on A-T splitting, this approach not only significantly enhances sensitivity and frequency resolution but also enables simultaneous extraction of amplitude and phase information from microwave signals. This provides new possibilities for the detection, reception, and demodulation of complex modulated signals\cite{ref2, ref3, ref19, ref20, ref21, ref22, ref23, ref24, ref25, ref26}. 

In previous studies on heterodyne Rydberg atom receivers, the static solution analysis method was commonly used\cite{ref3, ref27, ref28, ref29}. For a four-level system with only the local oscillator signal present, under steady-state conditions, the density matrix element $\rho_{ba}$ has a static solution independent of time, where $\Omega_m$ represents the microwave Rabi frequency exciting the two Rydberg levels. Upon introducing a weak microwave electric field, $\Omega_m$ consists of both the local oscillator microwave and the target microwave, i.e. $\Omega_m=\Omega_{L}+\Omega_s e^{i\delta_s t}$, (where ${{\Omega }_{L}}$ and ${{\Omega }_{s}}$ are the Rabi frequencies of the local oscillator and weak microwave electric field, respectively, and ${{\delta }_{s}}$ is the frequency difference between them). By directly substituting this into the ${{\rho }_{ba}}$ of the four-level system, the expression for the output optical power can be derived as follows\cite{ref3}:
\begin{equation}
P\left( t \right)=\overline{{{P}_{0}}}+\frac{\partial \overline{{{P}_{0}}}}{\partial {{\Omega }_{L}}}{{\Omega }_{s}}cos\left( {{\delta }_{s}}t+{{\phi }_{s}} \right).
\label{eq1}
\end{equation}
Here, $\overline{P_0}$ represents the average power of the probe light after passing through the atomic medium. The above expression indicates that the output signal of the four-level heterodyne receiver is proportional to the microwave electric field to be received, where ${\partial \overline{P_0}\left( \Omega_L \right)}/{\partial \Omega_L}$ represents the receiver's gain with respect to the target signal, and the target signal is automatically down-converted to $\delta_s$. In summary, the static solution method can qualitatively elucidate the receiving mechanism of the four-level heterodyne receiver. However, in the static solution, the gain coefficient ${\partial \overline{P_0}\left( \Omega_L \right)}/{\partial \Omega_L}$ is independent of $\delta_s$, i.e., it is unrelated to the baseband frequency of the received signal. This implies that the static solution cannot reflect the frequency response of the receiver.

Due to the difference between $\Omega_L$ and $\Omega_s$, after the rotating frame transformation, the element corresponding to $\Omega_m$ in the Hamiltonian still retains a time-dependent component $\Omega_s e^{i \delta_s t}$. Consequently, in the four-level heterodyne system, the density matrix element $\rho_{ba}$ continuously varies with time, which contradicts the static solution condition $d\rho_{ba}/dt=0$. This indicates that, in a heterodyne scenario, a strictly static solution does not exist. Moreover, the system bandwidth and other characteristics of the receiver are fundamentally determined by the time-varying behavior of the density matrix elements. Therefore, to accurately reveal the response characteristics of the four-level heterodyne receiver, it is essential to obtain the dynamic solutions of the density matrix elements under non-steady-state conditions. Since 2024, the dynamic solutions of Rydberg atom receivers have gradually become a research hotspot. The group of Yang at the Shanghai Institute of Optics and Fine Mechanics, CAS, investigated the dynamic response of Rydberg receivers under weak probe light conditions and derived analytical expressions for this mode \cite{2023High}. Meanwhile, the group of Tang at Xidian University and the group of C.L. Holloway at NIST developed numerical solutions for the dynamic response of Rydberg atom receivers using perturbation theory and Fourier series approximation, respectively \cite{Ren:24}\cite{2024Observation}. In 1993, W.M.Itano pointed out that the quantum projection noise of quantum sensors originates from the random distribution of population across energy levels\cite{Itano_1993}. However, these  methods fail to reveal the sources of $\rho_{ba}$noise and cannot provide further analysis of the $\rho_{ba}$ noise performance.

This study employs the frequency-domain solution method for the dynamic density matrix equations to derive the analytical dynamic solution of the receiver’s density matrix element $\rho_{ba}$, comprehensively accounting for the influence of thermal atoms during the solution process. The analytical dynamic solution shows that the receiver converts RF signals into baseband signals containing left and right sidebands, where the gain coefficients are $H_1\left(\omega \right)$ and $H_2\left(\omega \right)$, respectively. By leveraging the absorption characteristics of the medium, an expression relating the transmitted optical power to the received electric field can be further derived. This expression indicates that, for general modulated signals, the amplitude-frequency response of the receiver can be expressed as $\left| H_1(\omega) + H_2(\omega) \right|$, which is correlated with the baseband frequency of the target signal. This demonstrates that the dynamic solution provides a quantitative description of the receiver’s amplitude-frequency response, thereby enabling theoretical bandwidth estimation. Additionally, the dynamic solution is related to system parameters such as Rabi frequencies and level decay rates, making it possible to evaluate the impact of different parameters on the receiver’s bandwidth, thereby providing crucial theoretical guidance for system performance optimization. Based on the analytical expression of the density matrix element $\rho_{ba}$, it is demonstrated that the noise in $\rho_{ba}$ is primarily related to the population distribution probabilities across different energy levels, thus establishing the relationship between the theoretical ultimate sensitivity of the Rydberg atom receiver and the noise in $\rho_{ba}$ as well as the receiver's amplitude-frequency response. Finally, through experimental measurements and theoretical calculations, the amplitude-frequency response at different carrier frequencies is obtained, demonstrating that the bandwidth of the four-level heterodyne Rydberg atom receiver can exceed 10 MHz.

\section{Derivation of Dynamic Solution}
\begin{figure}[t]
    \centering
    \includegraphics[width=0.9\linewidth]{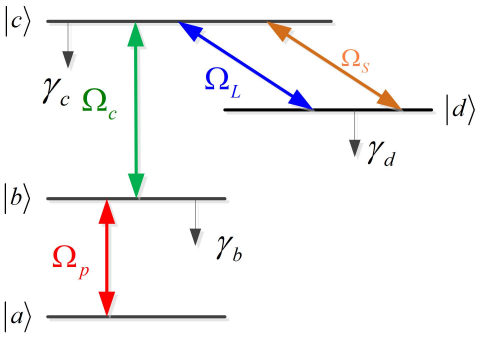}
    \caption{Energy level diagram of the Rydberg atom receiver.}
    \label{fig1}
\end{figure}

The energy level structure model of the Rydberg receiver is shown in Fig. \ref{fig1}. State $|a\rangle$ is the ground state, $|b\rangle$ is the intermediate state, and both $|c\rangle$ and $|d\rangle$ are Rydberg states. $\gamma_b$, $\gamma_c$, and $\gamma_d$ represent the decay rates of energy levels $b$, $c$, and $d$, respectively. The light coupled between energy levels $a$ and $b$ is referred to as the probe light, with a Rabi frequency of $\Omega_p$. The light coupled between energy levels $b$ and $c$ is called the coupling light, with a Rabi frequency of $\Omega_c$. There are two microwave electric fields coupled between energy levels $c$ and $d$: one is a strong local oscillator microwave electric field with a Rabi frequency of $\Omega_L$, and the other is a weak signal microwave electric field to be received, with a Rabi frequency of $\Omega_s$. The probe light, coupling light, and the local oscillator microwave electric field are perfectly resonant with the energy levels without any detuning, while the signal microwave electric field has a detuning of $\omega$ relative to the energy levels. The propagation direction of the probe light is along the z-axis, the coupling light propagates in the -z direction, and the microwave electric fields propagate along the x-axis.

Let the general modulated signal (amplitude/phase modulation) be represented as
\begin{equation}
    E_{s}\left( t \right)=A(t) cos \left( {{\omega }_{cd}}t+\varphi \left( t \right) \right).
\label{eq2}
\end{equation}
Here, $\omega_{cd}$ represents the carrier frequency of the target signal. Eq. (\ref{eq2}) can be rewritten in the following form:
\begin{equation}
    E_s(t) = \frac{\mathcal{E}(t)e^{i|\omega_c - \omega_d|t} + \mathcal{E}^*(t)e^{-i|\omega_c - \omega_d|t}}{2}.
\label{eq3}
\end{equation}
$|\omega_c - \omega_d|$ represents the resonant frequency between the $\left| c \right\rangle$ and $\left| d \right\rangle$ states, while $\mathcal{E}(t)$ encompasses both the modulation component of the signal and the detuning between the microwave field and the two Rydberg states, expressed as
\begin{equation}
    \mathcal{E}\left( t \right)=A\left( t \right){{e}^{i\left( {{\omega }_{cd}}-\left| {{\omega }_{c}}-{{\omega }_{d}} \right| \right)t}}{{e}^{i\varphi \left( t \right)}}.
\label{eq4}
\end{equation}
Assume that the probe light and coupling light propagate along the $+z$ and $-z$ directions, respectively, while the strong local oscillator field and the weak microwave field both propagate along the $+x$ direction. For the probe light, considering the position-dependent phase induced by electromagnetic wave propagation, the probe field can be expressed as
\begin{equation}
    {{E}_{opt}}\left( t \right)={{A}_{opt}}\cos \left( \left( {{\omega }_{b}}-{{\omega }_{a}} \right)t-{{k}_{p}}z \right).
\label{eq5}
\end{equation}
The coupling light and the microwave field can be described in a similar manner. At this point, the Hamiltonian matrix of the four-level heterodyne system is given by
\begin{subequations}
\begin{equation}
H = \frac{\hbar}{2}
    \begin{bmatrix}
        0 & \Omega_p e^{-i k_p z} & 0 & 0 \\
        \Omega_p e^{i k_p z} & 0 & \Omega_c e^{i k_c z} & 0 \\
        0 & \Omega_c e^{-i k_c z} & 0 & H(3,4) \\
        0 & 0 & H(4,3) & 0
    \end{bmatrix},
\label{eq6a}
\end{equation}
\begin{equation}
    H(3,4) = \Omega_L e^{-i k_m x} + \frac{e^{-i k_m x}}{2\pi} \int \Omega_s(\omega) e^{i \omega t} \, d\omega, 
\end{equation}
\label{eq6b}
\begin{equation}
    H(4,3) = \Omega_L e^{i k_m x} + \frac{e^{i k_m x}}{2\pi} \int \Omega_s^*(-\omega) e^{i \omega t} \, d\omega.
\label{eq6c}
\end{equation}
\label{eq6}
\end{subequations}

\noindent Here, $k_p$, $k_c$, and $k_m$ denote the wavevectors of the probe light, coupling light, and microwave field, respectively. Since the wavelengths of the weak microwave field and the local oscillator microwave field are approximately the same, they are both denoted as $k_m$. $\Omega_s (\omega)$ represents the Fourier transform of $\mathcal{E}(t)\cdot\wp_{cd}/\hbar$, $\wp_{cd}$ denotes the transition dipole moment, and $\hbar$ is the reduced Planck constant. The frequency coordinate $\omega$ represents the detuning between the received signal frequency and the microwave local oscillator electric field frequency. The frequency-domain variable $\omega$ in the Fourier transform of $\mathcal{E}(t)$ corresponds to the detuning between the received signal frequency and the resonance frequency of the Rydberg energy levels.

In the case of thermal atoms, the density matrix depends not only on time but also on spatial position. Under these conditions, the dynamical equation of the density matrix is given by\cite{sargent1974laser}
\begin{equation}
\left( {\frac{\partial }{{\partial t}} + {\bf{v}} \cdot \nabla } \right)\rho  =  - \frac{i}{\hbar }\left( {H\rho  - \rho H} \right) - \frac{1}{2}\left( {\Gamma \rho  + \rho \Gamma } \right) + \Lambda. 
\label{eq7}
\end{equation}

In Eq. (\ref{eq7}), $\mathbf{v}=\left( \begin{matrix}
   {{v}_{x}} & {{v}_{y}} & {{v}_{z}}  \\
\end{matrix} \right)$, $\nabla =\left( \begin{matrix}
   \frac{\partial }{\partial x} & \frac{\partial }{\partial y} & \frac{\partial }{\partial z}  \\
\end{matrix} \right)$. The non-zero elements of the \(4 \times 4\) relaxation matrix \(\Gamma\) and the \(4 \times 4\) repopulation matrix \(\Lambda\) for a four-level closed atom are:
$\Gamma _{1,1} = {\gamma _t}$,  $\Gamma _{2,2} = {\gamma _t} + {\gamma _b} + {\gamma _{dep\_b}}$,
$\Gamma_{3,3}= {\gamma _t} + {\gamma _c} + {\gamma _{dep\_c}}$,  $\Gamma_{4,4}= {\gamma _t} + {\gamma _d} + {\gamma _{dep\_d}}$.
${\Lambda _{1,1}} = \gamma _t^{}{\rho _{11}} + \left( {{\gamma _t} + {\gamma _b} + {\gamma _{dep\_b}}} \right){\rho _{22}} + \left( {{\gamma _t} + {\gamma _{dep\_c}}} \right){\rho _{33}} + \left( {{\gamma _t} + {\gamma _d} + {\gamma _{dep\_d}}} \right){\rho _{44}}$,
$\Lambda_{2,2}=\gamma_c\rho_{cc}$.
where \(\gamma_b, \gamma_c, \gamma_d\) are the decay rates of the levels \(|b\rangle, |c\rangle, |d\rangle\), respectively,  \(\gamma_t\) is the transit-time decay rate, and \(\gamma_{dep\_b}, \gamma_{dep\_c}, \gamma_{dep\_d}\) are the dephasing decay rates due to collisions and other factors. From this, six independent density matrix equations can be obtained: 
\begin{subequations}
\begin{equation}
\begin{cases}
\begin{aligned}
    2i &\left( \text{\small $\frac{\partial}{\partial t}$} + \mathbf{v} \cdot \nabla \right) \rho_{ba} = \Omega_p e^{i k_p z} (\rho_{aa} - \rho_{bb})  \\
    &+ \Omega_c e^{i k_c z} \rho_{ca}- i \gamma_{ba}^{\prime} \rho_{ba}, 
\end{aligned} \\
\begin{aligned}
    2i & \left( \text{\small $\frac{\partial}{\partial t}$} + \mathbf{v} \cdot \nabla \right) \rho_{ca} = \Omega_c e^{-i k_c z} \rho_{ba} + I_1 \rho_{da}  \\
    &- \Omega_p e^{i k_p z} \rho_{cb}- i \gamma_{ca}^{\prime} \rho_{ca},
\end{aligned} \\
\begin{aligned}
    2i & \left( \text{\small $\frac{\partial}{\partial t}$} + \mathbf{v} \cdot \nabla \right) \rho_{cb} = \Omega_c e^{-i k_c z} (\rho_{bb} - \rho_{cc}) \\
    & + I_1 \rho_{db} - \Omega_p e^{-i k_p z} \rho_{ca} - i \gamma_{cb}^{\prime} \rho_{cb},
\end{aligned} \\
\begin{aligned} 
    2i & \left( \text{\small $\frac{\partial}{\partial t}$} + \mathbf{v} \cdot \nabla \right) \rho_{da} = I_2 \rho_{ca} - \Omega_p e^{i k_p z} \rho_{db} \\
    & - i \gamma_{da}^{\prime}\rho_{da}, 
\end{aligned} \\
\begin{aligned} 
    2i & \left( \text{\small $\frac{\partial}{\partial t}$} + \mathbf{v} \cdot \nabla \right) \rho_{db} = I_2 \rho_{cb} - \Omega_p e^{-i k_p z} \rho_{da} \\
    & - \Omega_c e^{-i k_c z} \rho_{dc} - i \gamma_{db}^{\prime} \rho_{db},
\end{aligned} \\
\begin{aligned} 
    2i & \left( \text{\small $\frac{\partial}{\partial t}$} + \mathbf{v} \cdot \nabla \right) \rho_{dc} = I_2 (\rho_{cc} - \rho_{dd}) \\
    & - \Omega_c e^{i k_c z} \rho_{db} - i \gamma_{dc}^{\prime} \rho_{dc}.
\end{aligned}
\end{cases}
\label{eq8a}
\end{equation}

\begin{equation}
I_1 = \Omega_L e^{-i k_m x} + \frac{e^{-i k_m x}}{2\pi} \int \Omega_s(\omega) e^{i \omega t} \, d\omega,
\label{eq8b}
\end{equation}

\begin{equation}
I_2 = \Omega_L e^{i k_m x} + \frac{e^{i k_m x}}{2\pi} \int \Omega_s^*(-\omega) e^{i \omega t} \, d\omega.
\label{eq8c}
\end{equation}
\label{eq8}
\end{subequations}
where $\gamma _{ij}^{\prime} = \gamma _i^{\prime} + \gamma_j^{\prime}$, $\gamma_a^{\prime}=\gamma_t$, $\gamma_b^{\prime}=\gamma_b+\gamma_t+\gamma_{dep\_b}$, $\gamma_c^{\prime}=\gamma_c+\gamma_t+\gamma_{dep\_c}$,
$\gamma_d^{\prime}=\gamma_d+\gamma_t+\gamma_{dep\_d}$.

Solving the system of equations (\ref{eq8}) directly is too complex, so it can be simplified by using an approximation. The diagonal elements of the density matrix, $\rho_{ii}( i=a,b,c,d)$, represent the probability distribution of particles in each energy level. When the signal field is relatively weak, $\rho_{ii}$ can be approximated as the population probability $\rho_{ii}^0$ without the signal field. For the off-diagonal elements $\rho_{ij}(i,j=a,b,c,d,i \ne j)$, the product term involving the time-varying electric field appears in Eq. (\ref{eq8}):
\begin{equation}
    \rho_{ij}\left(t\right)\frac{e^{-ik_mx}}{2\pi}\int\Omega_s^*\left(-\omega\right)e^{i\omega t}d\omega.
\label{eq9}
\end{equation}
The presence of the product of two time-varying terms in Eq. (\ref{eq9}) makes the system of differential equations difficult to solve. To address this, $\rho_{ij}$ is expressed as:$ {{\rho }_{ij}}(t)=\rho_{ij}^{\widetilde{0}}+e(t)$. $\rho_{ij}^{\widetilde{0}}$ represents the density matrix element of the system in the absence of the weak microwave field, and $e(t)$ denotes a small perturbation of $\rho_{ij}$ induced by the presence of the signal field, which is related to $\mathcal{E}(t)$ in Eq. (\ref{eq3}). Term $\rho _{ij}^{\widetilde{0}}\int{\Omega _{s}^{*}(-\omega){{e}^{i\omega t}}d\omega }$ corresponds to the part of the output signal that varies linearly with $\mathcal{E}(t)$, while term $e(t)\int{\Omega _{s}^{*}(-\omega){{e}^{i\omega t}}d\omega }$ represents higher-order terms of $\mathcal{E}(t)$. Under weak field conditions, the relation \text{\small $\left| \rho _{ij}^{\widetilde{0}} \right|\gg \left| e\left( t \right) \right|$} holds, indicating that the nonlinear terms in the four-level heterodyne system are much smaller than the linear terms. This behavior is consistent with the general requirements for a standard receiver. Therefore, Eq. (\ref{eq9}) can be approximated as:
\begin{equation}
    \rho _{ij}^{\widetilde{0}}\frac{{{e}^{-i{{k}_{m}}x}}}{2\pi }\int{\Omega _{s}^{*}\left( -\omega  \right)}{{e}^{j\omega t}}d\omega.
\label{eq10}
\end{equation}

In addition, the laser beam diameter is approximately 1 mm, which is much smaller than the microwave wavelength. Given that $k_m=2\pi/\lambda_m$, it follows that $e^{\pm ik_m x}\approx1$. After the above approximations, Eq. (\ref{eq8}) reduces to a set of first-order linear differential equations with constant coefficients:
\begin{equation}
\begin{cases}
\begin{aligned}
& 2i\left(\text{\small $\frac{\partial}{\partial t}$}+\mathbf{v}\cdot\nabla\right)\rho_{ba} = \Omega_{p}e^{ik_{p}z}\left(\rho_{aa}^{0}-\rho_{bb}^{0}\right)  \\
& \quad +\Omega_{c}e^{ik_{c}z}\rho_{ca}-i\gamma_{ba}^{\prime}\rho_{ba}, \\
& 2i\left(\text{\small $\frac{\partial}{\partial t}$}+\mathbf{v}\cdot\nabla\right)\rho_{ca} = \Omega_ce^{-ik_cz}\rho_{ba}+\Omega_L\rho_{da}  \\
& \quad +\frac{\rho_{da}^{\widetilde{0}}}{2\pi}\int\Omega_s\left(\omega\right)e^{i\omega t}d\omega-\Omega_pe^{ik_pz}\rho_{cb}-i\gamma_{ca}^{\prime}\rho_{ca}, \\
& 2i\left(\text{\small $\frac{\partial}{\partial t}$}+\mathbf{v}\cdot\nabla\right)\rho_{cb} = \Omega_ce^{-ik_cz}\left(\rho_{bb}^0-\rho_{cc}^0\right)+\Omega_L\rho_{db}  \\
& \quad +\frac{\rho_{db}^{\widetilde{0}}}{2\pi}\int\Omega_s\left(\omega\right)e^{i\omega t}d\omega-\Omega_pe^{-ik_pz}\rho_{ca}-i\gamma_{cb}^{\prime}\rho_{cb}, \\
& 2i\left(\text{\small $\frac{\partial}{\partial t}$}+\mathbf{v}\cdot\nabla\right)\rho_{da} = \frac{\rho_{ca}^{\widetilde{0}}}{2\pi}\int\Omega_s^*\left(-\omega\right)e^{i\omega t}d\omega  \\
& \quad -\Omega_pe^{ik_pz}\rho_{db}+\Omega_L\rho_{ca}-i\gamma_{da}^{\prime}\rho_{da}, \\
& 2i\left(\text{\small $\frac{\partial}{\partial t}$}+\mathbf{v}\cdot\nabla\right)\rho_{db} = \Omega_{L}\rho_{cb}+\frac{\rho_{cb}^{\widetilde{0}}}{2\pi}\int\Omega_{s}^{*}\left(-\omega\right)e^{i\omega t}d\omega  \\
& \quad -\Omega_{p}e^{-ik_{p}z}\rho_{da}-\Omega_{c}e^{-ik_{c}z}\rho_{dc}-i\gamma_{db}^{\prime}\rho_{db}, \\
& 2i\left(\text{\small $\frac{\partial}{\partial t}$}+\mathbf{v}\cdot\nabla\right)\rho_{dc} = \Omega_L\left(\rho_{cc}^0-\rho_{dd}^0\right)-\Omega_ce^{ik_cz}\rho_{db}  \\
& \quad +\frac{\rho_{cc}^0-\rho_{dd}^0}{2\pi}\int\Omega_s^*\left(-\omega\right)e^{i\omega t}d\omega-i\gamma_{dc}^{\prime}\rho_{dc}.
\end{aligned}
\end{cases}
\label{eq11}
\end{equation}

The diagonal elements of the density matrix, $\rho_{ii}$, are real and independent of the spatial coordinate $z$. The off-diagonal elements, $\rho_{ij}^{\tilde{0}}$, can be factorized into two components: a $z$-independent term, $\rho_{ij}^0$ (see Appendix A for details), and a position-dependent phase factor (derived in Appendix B), i.e., 
\begin{equation}
\begin{aligned}
 & \rho_{ba}^{\widetilde{0}}=\rho_{ba}^0e^{ik_pz},\rho_{ca}^{\widetilde{0}}=\rho_{ca}^0e^{-i\left(k_c-k_p\right)z}, \\
 & \rho_{da}^{\widetilde{0}}=\rho_{da}^0e^{-i\left(k_c-k_p\right)z}, \rho_{cb}^{\widetilde{0}}=\rho_{cb}^0e^{-ik_cz}, \\
 & \rho_{db}^{\widetilde{0}}=\rho_{db}^0e^{-ik_cz},\rho_{dc}^{\widetilde{0}}=\rho_{dc}^0. 
\end{aligned}
\label{eq12}
\end{equation}
Since the microwave field propagates along the $x$ direction and satisfies $e^{\pm ik_mx}\approx1$, the position-dependent phase factor in $\rho _{dc}^{\widetilde{0}}$ can be neglected. Substituting Eq. (\ref{eq12}) into Eq. (\ref{eq11}) further simplifies the expression. 

Next, we aim to derive the relationship between $\rho_{ba}$ and the signal fields $\mathcal{E}(t)$ and $\mathcal{E}^*(t)$, as the atomic receiver detects the incoming signal through the absorption characteristics of the probe light within the atomic medium. This relationship is governed by the system of differential equations given in (\ref{eq11}). However, directly solving this multivariable linear differential system remains very challenging. The system described by these differential equations is a linear time-invariant (LTI) system, whose complete response consists of a zero-input response and a zero-state response. The zero-input response is determined solely by the initial conditions of the system, while the zero-state response is determined entirely by the external input. Over time, the influence of the initial conditions decays to zero, and the system output becomes fully determined by the input signal. To obtain the zero-state response of the Rydberg atomic receiver, we can solve the system of differential equations (\ref{eq11}) by applying a Fourier transform. This approach converts the time-domain differential equations into algebraic equations in the frequency domain, thereby simplifying the solution process. Through frequency-domain analysis, the zero-state response of the system can be explicitly obtained, offering clearer insights into its response characteristics.

The terms on the left-hand side of the equations in Eq. (\ref{eq11}) depend on the spatial coordinates $x$, $y$, and $z$, while the terms on the right-hand side depend only on time $t$ and the $z$ coordinate. Therefore, it is sufficient to perform a two-dimensional Fourier transform with respect to $t$ and $z$. The frequency-domain expression of the density matrix elements is denoted as $\rho_{ij}(\omega,\omega_z)$. The results of the Fourier transform of Eq. (\ref{eq11}) are shown in Eq. (\ref{eq13}). The final solution yields the frequency-domain expression of ${{\rho }_{ba}}$ as shown in Eq. (\ref{eq14a}), where:
\begin{figure*}[!ht]
\begin{subequations}
\begin{align}
 & \begin{aligned}
    \left( i{{\gamma }_{ba}^{\prime}}-2\left( \omega +{{v}_{z}}{{\omega }_{z}} \right) \right){{\rho }_{ba}}\left( \omega ,{{\omega }_{z}} \right)={{\left( 2\pi  \right)}^{2}}{{\Omega }_{p}}\left( \rho _{aa}^{0}-\rho _{bb}^{0} \right)\delta \left( \omega  \right)\delta \left( {{\omega }_{z}}-{{k}_{p}} \right)+{{\Omega }_{c}}{{\rho }_{ca}}\left( \omega ,{{\omega }_{z}}-{{k}_{c}} \right),
 \end{aligned}
 \label{eq13a} \\
 & \begin{aligned}
    \left( i{{\gamma }_{ca}^{\prime}}-2\left( \omega +{{v}_{z}}{{\omega }_{z}} \right) \right){{\rho }_{ca}}\left( \omega ,{{\omega }_{z}} \right) 
    & ={{\Omega }_{c}}{{\rho }_{ba}}\left( \omega ,{{\omega }_{z}}+{{k}_{c}} \right)+{{\Omega }_{L}}{{\rho }_{da}}\left( \omega ,{{\omega }_{z}} \right)+2\pi \rho _{da}^{0}{{\Omega }_{s}}\left( \omega  \right)\delta \left( {{\omega }_{z}}+{{k}_{c}}-{{k}_{p}} \right) \\
    & -{{\Omega }_{p}}{{\rho }_{cb}}\left( \omega ,{{\omega }_{z}}-{{k}_{p}} \right), 
 \end{aligned} 
 \label{eq13b} \\
 & \begin{aligned}
    \left( i{{\gamma }_{cb}^{\prime}}-2\left( \omega +{{v}_{z}}{{\omega }_{z}} \right) \right){{\rho }_{cb}}\left( \omega ,{{\omega }_{z}} \right)
    & ={{\left( 2\pi  \right)}^{2}}{{\Omega }_{c}}\left( \rho _{bb}^{0}-\rho _{cc}^{0} \right)\delta \left( \omega  \right)\delta \left( {{\omega }_{z}}+{{k}_{c}} \right)+{{\Omega }_{L}}{{\rho }_{db}}\left( \omega ,{{\omega }_{z}} \right) \\
    & +2\pi \rho _{db}^{0}{{\Omega }_{s}}\left( \omega  \right)\delta \left( {{\omega }_{z}}+{{k}_{c}} \right)-{{\Omega }_{p}}{{\rho }_{ca}}\left( \omega ,{{\omega }_{z}}+{{k}_{p}} \right), 
    \end{aligned} 
    \label{eq13c} \\
 & \begin{aligned}
    \left( i{{\gamma }_{da}^{\prime}}-2\left( \omega +{{v}_{z}}{{\omega }_{z}} \right) \right){{\rho }_{da}}\left( \omega ,{{\omega }_{z}} \right)={{\Omega }_{L}}{{\rho }_{ca}}\left( \omega ,{{\omega }_{z}} \right)+2\pi \rho _{ca}^{0}\Omega _{s}^{*}\left( -\omega  \right)\delta \left( {{\omega }_{z}}+{{k}_{c}}-{{k}_{p}} \right)-{{\Omega }_{p}}{{\rho }_{db}}\left( \omega ,{{\omega }_{z}}-{{k}_{p}} \right), 
    \end{aligned} 
    \label{eq13d} \\
 & \begin{aligned}
    \left( i{{\gamma }_{db}^{\prime}}-2\left( \omega +{{v}_{z}}{{\omega }_{z}} \right) \right){{\rho }_{db}}\left( \omega ,{{\omega }_{z}} \right)
    & ={{\Omega }_{L}}{{\rho }_{cb}}\left( \omega ,{{\omega }_{z}} \right)+2\pi \rho _{cb}^{0}\Omega _{s}^{*}\left( -\omega  \right)\delta \left( {{\omega }_{z}}+{{k}_{c}} \right)-{{\Omega }_{p}}{{\rho }_{da}}\left( \omega ,{{\omega }_{z}}+{{k}_{p}} \right) \\
    & -{{\Omega }_{c}}{{\rho }_{dc}}\left( \omega ,{{\omega }_{z}}+{{k}_{c}} \right), 
 \end{aligned} 
 \label{eq13e} \\
 & \begin{aligned}
    \left( i{{\gamma }_{dc}^{\prime}}-2\left( \omega +{{v}_{z}}{{\omega }_{z}} \right) \right){{\rho }_{dc}}\left( \omega ,{{\omega }_{z}} \right)
    & ={{\left( 2\pi  \right)}^{2}}{{\Omega }_{L}}\left( \rho _{cc}^{0}-\rho _{dd}^{0} \right)\delta \left( \omega  \right)\delta \left( {{\omega }_{z}} \right)+2\pi \left( \rho _{cc}^{0}-\rho _{dd}^{0} \right)\Omega _{s}^{*}\left( -\omega  \right)\delta \left( {{\omega }_{z}} \right) \\
    & -{{\Omega }_{c}}{{\rho }_{db}}\left( \omega ,{{\omega }_{z}}-{{k}_{c}} \right).
 \end{aligned}
 \label{eq13f}
\end{align}
\label{eq13}
\end{subequations}
\hrulefill
\end{figure*}

\begin{figure*}[!hbt]
\begin{subequations}
\begin{equation}
\rho_{ba}\left(\omega,\omega_z\right) =\left(2\pi\right)^2\rho_{ba}^0\delta\left(\omega\right)\delta\left(\omega_z-k_p\right)+2\pi H_{1v}(\omega)\Omega_s(\omega)\delta{\left(\omega_z-k_p\right)}+2\pi H_{2v}(\omega)\Omega_s^*(-\omega)\delta{\left(\omega_z-k_p\right)}.
\label{eq14a}
\end{equation}

\begin{equation}
\rho _{ba}^{0}=\frac{{{\Omega }_{p}}}{\left( i{{\gamma }_{ba}^{\prime}}-2{{v}_{z}}{{k}_{p}}-\frac{{{\Omega }_{c}}^{2}}{E\left( 0,{{k}_{p}}-{{k}_{c}} \right)} \right)}\left( \begin{aligned}
  & \left( \rho _{aa}^{0}-\rho _{bb}^{0} \right)-\frac{{{\Omega }_{c}}}{E\left( 0,{{k}_{p}}-{{k}_{c}} \right)} \\ 
 & \left( \begin{aligned}
  & {{C}_{0}}+\frac{{{\Omega }_{L}}^{2}}{A\left( 0,-{{k}_{c}} \right)D\left( 0,{{k}_{p}}-{{k}_{c}} \right)} \\ 
 & \left( {{C}_{0}}-\frac{{{\Omega }_{c}}\left( \rho _{cc}^{0}-\rho _{dd}^{0} \right)}{i{{\gamma }_{dc}^{\prime}}} \right)\left( 1+\frac{{{\Omega }_{p}}^{2}}{A\left( 0,-{{k}_{c}} \right)B\left( 0,-{{k}_{c}} \right)} \right) \\ 
\end{aligned} \right) \\ 
\end{aligned} \right),
\label{eq14b}
\end{equation}

\begin{equation}
  {{H}_{1v}}=\frac{{{\Omega }_{c}}}{\left( i{{\gamma }_{ba}^{\prime}}-2\left( \omega +{{k}_{p}}{{v}_{z}} \right)-\frac{{{\Omega }_{c}}^{2}}{E\left( \omega ,{{k}_{p}}-{{k}_{c}} \right)} \right)E\left( \omega ,{{k}_{p}}-{{k}_{c}} \right)}\left( \begin{aligned}
  & \rho _{da}^{0}-\frac{{{\Omega }_{p}}\rho _{db}^{0}}{B\left( \omega ,-{{k}_{c}} \right)} \\ 
 & \left( \begin{aligned}
  & 1+\frac{{{\Omega }_{L}}^{2}}{A\left( \omega ,-{{k}_{c}} \right)D\left( \omega ,{{k}_{p}}-{{k}_{c}} \right)} \\ 
 & \left( 1+\frac{{{\Omega }_{p}}^{2}}{A\left( \omega ,-{{k}_{c}} \right)B\left( \omega ,-{{k}_{c}} \right)} \right) \\ 
\end{aligned} \right) \\ 
\end{aligned} \right),
\label{eq14c}
\end{equation}

\begin{equation}
{{H}_{2v}}={{\Omega }_{L}}{{\Omega }_{c}}\frac{\left( \begin{aligned}
  & \frac{1}{D\left( \omega ,{{k}_{p}}-{{k}_{c}} \right)}\left( 1+\frac{{{\Omega }_{p}}^{2}}{A\left( \omega ,-{{k}_{c}} \right)B\left( \omega ,-{{k}_{c}} \right)} \right)\rho _{ca}^{0}-\frac{{{\Omega }_{p}}}{A\left( \omega ,-{{k}_{c}} \right)}\left( \rho _{cb}^{0}-\frac{{{\Omega }_{c}}\left( \rho _{cc}^{0}-\rho _{dd}^{0} \right)}{\left( i{{\gamma }_{dc}^{\prime}}-2\omega  \right)} \right) \\ 
 & \left( \frac{1}{B\left( \omega ,-{{k}_{c}} \right)}+\frac{1}{D\left( \omega ,{{k}_{p}}-{{k}_{c}} \right)}\left( 1+\frac{{{\Omega }_{p}}^{2}}{A\left( \omega ,-{{k}_{c}} \right)B\left( \omega ,-{{k}_{c}} \right)} \right)\left( 1+\frac{{{\Omega }_{L}}^{2}}{A\left( \omega ,-{{k}_{c}} \right)B\left( \omega ,-{{k}_{c}} \right)} \right) \right) \\ 
\end{aligned} \right)}{\left( i{{\gamma }_{ba}^{\prime}}-2\left( \omega +{{k}_{p}}{{v}_{z}} \right)-\frac{{{\Omega }_{c}}^{2}}{E\left( \omega ,{{k}_{p}}-{{k}_{c}} \right)} \right)E\left( \omega ,{{k}_{p}}-{{k}_{c}} \right)}.
\label{eq14d}
\end{equation}
\label{eq14}
\end{subequations}
\hrulefill
\end{figure*}

\begin{subequations}
\begin{equation}
\begin{aligned}
    A\left(\omega,\omega_{z}\right) & =i\gamma_{db}^{\prime}-2\left(\omega+v_z\omega_z\right) \\
    & -\frac{\Omega_c^2}{\left(i\gamma_{dc}^{\prime}-2\left(\omega+v_z\left(\omega_z+k_c\right)\right)\right)},
\end{aligned} 
\label{eq15a}
\end{equation}

\begin{equation}
  B\left( \omega ,{{\omega }_{z}} \right)= i{{\gamma }_{cb}^{\prime}}-2\left( \omega +{{v}_{z}}{{\omega }_{z}} \right)-\frac{{{\Omega }_{L}}^{2}}{A\left( \omega, 
  {{\omega }_{z}} \right)},
\label{eq15b}
\end{equation}

\begin{equation}
  {{C}_{0}}=\frac{{{\Omega }_{c}}}{B\left( 0,-{{k}_{c}} \right)}\left( \left( \rho _{bb}^{0}-\rho _{cc}^{0} \right)-\frac{{{\Omega }_{L}}^{2}\left( \rho 
  _{cc}^{0}-\rho _{dd}^{0} \right)}{i{{\gamma }_{dc}^{\prime}}A\left( 0,-{{k}_{c}} \right)} \right),
\label{eq15c}
\end{equation}

\begin{equation}
\begin{aligned} 
   D\left( \omega ,{{\omega }_{z}} \right) &=
   i{{\gamma }_{da}^{\prime}}-2\left( \omega +{{v}_{z}}{{\omega }_{z}} \right)-\frac{{{\Omega }_{p}}^{2}}{A\left( \omega ,{{\omega }_{z}}-{{k}_{p}} \right)} \\ 
   & \times \left( 1+\frac{{{\Omega }_{L}}^{2}}{A\left( \omega ,{{\omega }_{z}}-{{k}_{p}} \right)B\left( \omega ,{{\omega }_{z}}-{{k}_{p}} \right)} \right),
\end{aligned}
\label{eq15d}
\end{equation}

\begin{equation}
\begin{aligned}
  & E\left( \omega ,{{\omega }_{z}} \right)=i{{\gamma }_{ca}^{\prime}}-2\left( \omega +{{v}_{z}}{{\omega }_{z}} \right)-\frac{{{\Omega }_{p}}^{2}}{B\left( \omega ,{{\omega }_{z}}-{{k}_{p}} \right)} \\ 
  & -\frac{{{\Omega }_{L}}^{2}}{D\left( \omega ,{{\omega }_{z}} \right)}{{\left( 1+\frac{{{\Omega }_{p}}^{2}}{A\left( \omega ,{{\omega }_{z}}-{{k}_{p}} \right)B\left( \omega ,{{\omega }_{z}}-{{k}_{p}} \right)} \right)}^{2}}.
\end{aligned}
\label{eq15e}
\end{equation}
\end{subequations}

Under weak probe light conditions, most of the particles remain in the ground state, i.e., $\rho _{aa}^{0}\approx 1$, $\rho _{bb}^{0}\approx 0$, ${{\rho }_{cb}}\approx 0$, and ${{\rho }_{db}}\approx 0$. In this case, the dynamic solution of the system can be obtained using only Eqs. (\ref{eq13a}), (\ref{eq13b}), and (\ref{eq13d}). The calculated results are as follows:
\begin{equation}
\begin{aligned}
    \rho_{ba}\left(\omega,\omega_z\right) & =\left(2\pi\right)^2\rho_{ba}^0\delta\left(\omega\right)\delta\left(\omega_z-k_p\right) \\
    & +2\pi H_{1v}(\omega)\Omega_s(\omega)\delta{\left(\omega_z-k_p\right)}\\
    & +2\pi H_{2v}(\omega)\Omega_s^*(-\omega)\delta{\left(\omega_z-k_p\right)}.
\end{aligned}
\label{eq16}
\end{equation}
Eq. (\ref{eq16}) has the same form as Eq. (\ref{eq14a}), where:

\begin{subequations}
\begin{equation}
    \text{\small $\rho _{ba}^{0}=\frac{{{\Omega }_{p}}}{{{B}_{w}}\left( 0,{{k}_{p}} \right)}$},
\label{eq17a}
\end{equation}
\begin{equation}
    \text{\small ${{H}_{1v}}\left( \omega  \right)=\frac{{{\Omega }_{p}}\Omega _{c}^{2}{{\Omega }_{L}}}{{{C}_{w}}\left( \omega  \right){{C}_{w}}\left( 0 \right)\left( i{{\gamma }_{da}^{\prime}}-2\left( \left( {{k}_{p}}-{{k}_{c}} \right){{v}_{z}} \right) \right)}$},
\label{eq17b}
\end{equation}
\begin{equation}
    \text{\small ${{H}_{2v}}\left( \omega  \right)=\frac{{{\Omega }_{p}}\Omega _{c}^{2}{{\Omega }_{L}}}{{{C}_{w}}\left( \omega  \right){{C}_{w}}\left( 0 \right)\left( i{{\gamma }_{da}^{\prime}}-2\left( \omega +\left( {{k}_{p}}-{{k}_{c}} \right){{v}_{z}} \right) \right)}$},
\label{eq17c}
\end{equation}
\begin{equation}
    \text{\small ${{A}_{w}}\left( \omega ,{{\omega }_{z}} \right)=i{{\gamma }_{ca}^{\prime}}-2\left( \omega +{{\omega }_{z}}{{v}_{z}} \right)-\frac{\Omega _{L}^{2}}{\left( i{{\gamma }_{da}^{\prime}}-2\left( \omega +{{\omega }_{z}}{{v}_{z}} \right) \right)}$},
\label{eq17d}
\end{equation}
\begin{equation}
    \text{\small ${{B}_{w}}\left( \omega ,{{\omega }_{z}} \right)=i{{\gamma }_{ba}^{\prime}}-2\left( \omega +{{\omega }_{z}}{{v}_{z}} \right)-\frac{\Omega _{c}^{2}}{{{A}_{w}}\left( \omega ,{{\omega }_{z}}-{{k}_{c}} \right)}$},
\label{eq17e}
\end{equation}
\begin{equation}
    \text{\small ${{C}_{w}}\left( \omega  \right)={{B}_{w}}\left( \omega ,{{k}_{p}} \right){{A}_{w}}\left( \omega ,{{k}_{p}}-{{k}_{c}} \right)$}.
\label{eq17f}
\end{equation}
\label{eq17}
\end{subequations}

The weak probe approximation provides a simplified approach for solving the dynamical equations; however, to ensure accuracy, the subsequent calculations are carried out using the full, non-approximated dynamical solution. Finally, the time-domain expressions can be obtained by performing inverse Fourier transforms on Eq. (\ref{eq14a}):
\begin{equation}
\begin{aligned}
    \rho_{ba}\left(t,z,v_z\right)&=\rho_{ba}^0e^{ik_pz}+\frac{\wp_{cd}}{\hbar}h_{1v}\left(t\right)\otimes\mathcal{E}\left(t\right)e^{ik_pz}\\
    &+\frac{\wp_{cd}}{\hbar}h_{2v}\left(t\right)\otimes\mathcal{E}^*\left(t\right)e^{ik_pz}.
\end{aligned}
\label{eq18}
\end{equation}

Eq. (\ref{eq14a}) and Eq. (\ref{eq18}) show that the density matrix element comprises two components. The first corresponds to the first term and determines the average transmission power of the probe light after traversing the atomic vapor cell. The second, consisting of the remaining terms, encodes the microwave signal information. Notably, the frequency of the signal $\mathcal{E}(t)$ equals the difference frequency between $E_s(t)$ and the local oscillator microwave electric field, illustrating the heterodyne receiver’s intrinsic down-conversion mechanism. Furthermore, the equations reveal that the relationship between the output signal and the input microwave field is not simply linear. In the following section, we derive the system response function, incorporating the effects of atomic thermal motion and various modulation conditions.

\section{Analysis on the Dynamic Solution Response}
\begin{figure*}[!ht]
    \centering
    \includegraphics[width=0.65\linewidth]{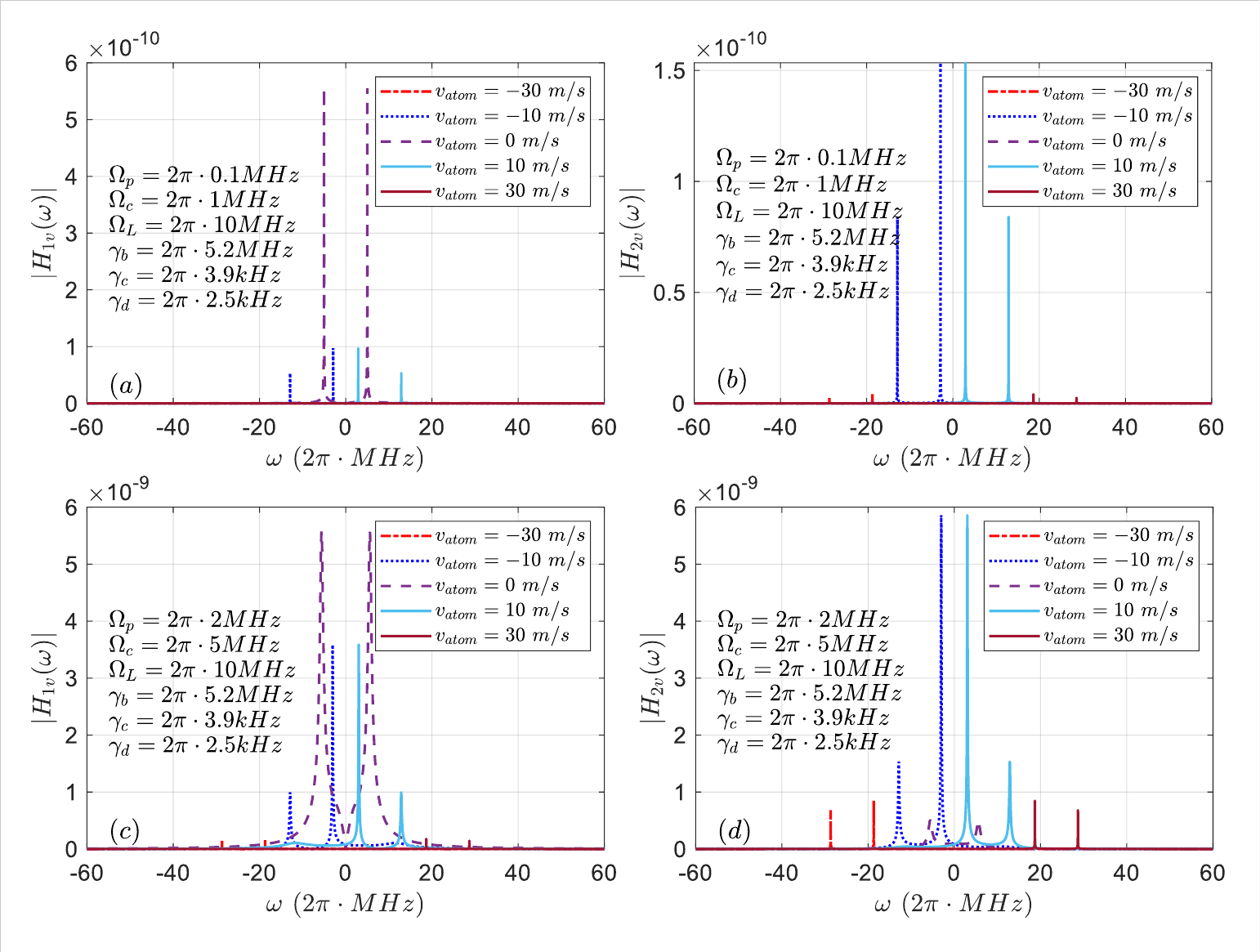}
    \caption{Response of atoms with different velocities to the microwave field. (a) and (b) present the amplitude response curves of $H_{1v}(\omega)$ and $H_{2v}(\omega)$, respectively, under low Rabi frequencies of the probe and coupling fields. (c) and (d) show the corresponding amplitude response curves under high Rabi frequencies of the probe and coupling fields.}
    \label{fig2}
\end{figure*}

From Eq. (\ref{eq18}), it is evident that the density matrix element $\rho_{ba}$ varies with different atomic velocities $v_z$. Figs. \ref{fig2}(a) and \ref{fig2}(c) depict the amplitude response $H_{1v}(\omega)$ for atoms at various velocities under different parameter conditions, while Figs. \ref{fig2}(b) and \ref{fig2}(d) display the amplitude response $H_{2v}(\omega)$ under the same conditions. It is observed that the peak value of the zero-velocity curve in Figs. \ref{fig2}(a) and \ref{fig2}(c) is significantly higher than that in Figs. \ref{fig2}(b) and \ref{fig2}(d), indicating that the response of low-velocity atoms near zero velocity to the microwave field is primarily governed by $H_{1v}(\omega)$. Furthermore, when the atomic velocity is zero, the heterodyne Rydberg receiver exhibits a maximum response amplitude at frequencies of $\pm \Omega_L/2$, attributed to the A–T splitting effect. For nonzero atomic velocities, the induced Doppler shift results in a lateral displacement of the amplitude response curve. Additionally, when the Rabi frequencies of the probe and coupling are relatively low, the amplitude response curve displays two distinct sharp peaks; conversely, with higher Rabi frequencies, the response curve becomes broadened.

As shown in Fig. \ref{fig2}, under a single velocity condition, the atomic response to the microwave electric field manifests as impulses at two specific frequency points, indicating that atoms at this velocity respond only to those two frequencies. However, the atomic vapor cell contains atoms with varying velocities, which collectively enable the cell to respond to microwave signals within a certain bandwidth range. Moreover, atoms with different velocities contribute differently to the amplitude response of the receiver: atoms with higher velocities exhibit a weaker response. Since the velocity distribution of atoms in the vapor cell follows the Maxwell–Boltzmann distribution, the susceptibility $\chi$ of the heterodyne four-level system can be expressed as a Doppler average of $\chi_v(v_z)$: 
\begin{equation}
    \chi=\int N(v_z)\chi_v(v_z)dv_z,
\label{eq19}
\end{equation}
where $N(v_z)=N_0 f(v_z)$, with $N_0$ denoting the atomic density and $f(v_z)$ representing the Maxwell–Boltzmann distribution. In the calculation of $\chi$, we set the effective integration interval to $ [-3\sigma(v_z) ,3\sigma(v_z)]$, divide it into numerous small subintervals, and employ a numerical summation approach to compute the integral.

The susceptibility $\chi_v(v_z)$ is given by\cite{sargent1974laser}:
\begin{subequations}
\begin{align}
    &{{\chi }_{v}}\left( {{v}_{z}} \right)=-\frac{2{{\wp }_{ab}}}{\mathcal{N}{{\varepsilon }_{0}}{{E}_{ab}}}\int\limits_{0}^{L}{{{\rho }_{ba}}\left( t,z,{{v}_{z}} \right)U^{*}\left( z \right)dz}, \\
    &\mathcal{N}=\int\limits_{0}^{L}{{{\left| U\left( z \right) \right|}^{2}}}dz,\ U\left( z \right)={{e}^{i{{k}_{p}}z}}.
\end{align} 
\label{eq20}
\end{subequations}

\noindent Where $\varepsilon_0$ is the vacuum permittivity, $E_{ab}$ is the electric field amplitude of the probe light, and $\Omega_p=E_{ab}\wp/\hbar$. By combining Eqs. (18), (19), and (20), the susceptibility of thermal atoms to the microwave field can be obtained as:
\begin{equation}
    \chi \left( t \right)=-\frac{2{{N}_{0}}\wp _{ab}^{2}}{{{\varepsilon }_{0}}\hbar {{\Omega }_{p}}}
    \left(
    \begin{aligned}
        & \overline{\rho _{ba}^{0}}+\frac{{{\wp }_{cd}}}{\hbar }{{h}_{1}}\left( t \right)\otimes \mathcal{E}\left( t \right) \\
        &+\frac{{{\wp }_{cd}}}{\hbar }{{h}_{2}}\left( t \right)\otimes {{\mathcal{E}}^{*}}\left( t \right)
    \end{aligned}
    \right),
\label{eq21}
\end{equation}
where $\overline{\rho _{ba}^{0}}=\int{\rho _{ba}^{0}}f\left( {{v}_{z}} \right)d{{v}_{z}}$, ${{h}_{1}}\left( t \right)=\int{{{h}_{1v}}\left( t \right)}f\left( {{v}_{z}} \right)d{{v}_{z}}$, ${{h}_{2}}\left( t \right)=\int{{{h}_{2v}}\left( t \right)}f\left( {{v}_{z}} \right)d{{v}_{z}}$, and "$\otimes$" denotes convolution operation. Eq. (\ref{eq21}) indicates that the susceptibility of the medium can be divided into two parts. The static component is independent of the microwave field and only depends on the Rabi frequencies of the probe, coupling, and local oscillator fields that excite the four-level system, representing the average transmitted power of the probe light through the medium. The time-varying component is related to the weak microwave field. This part consists of the sum of the convolutions of the input field $\mathcal{E}(t)$ and its complex conjugate $\mathcal{E}^*(t)$ with the response functions $h_1(t)$ and $h_2(t)$, respectively. Assuming that $\mathcal{E}(t)$ is a single-frequency signal, after passing through the atomic medium, it generates two sidebands at positive and negative frequencies with unequal powers, as described by Eq. (\ref{eq21}). The static solution cannot accurately capture this asymmetry between the positive and negative sidebands. Meanwhile, the response functions $h_1(t)$ and $h_2(t)$ in Eq. (\ref{eq21}) reflect the frequency-domain characteristics of the density matrix element. Compared to the static solution, the dynamic solution better reflects the intrinsic response of the four-level heterodyne Rydberg atom receiver to microwaves, namely that the system response is inherently tied to the detuning frequency between the weak microwave field and the Rydberg levels. Eq. (\ref{eq21}) captures this relationship.

The bandwidth  of the Rydberg atom receiver should be determined jointly by \( h_1(t) \) and \( h_2(t) \). Assuming the received signal is a single-frequency signal \( \mathcal{E}(t) = e^{i\omega t} \), then the power of the probe light transmitted through the atomic vapor cell is:

\begin{equation}
\begin{array}{l}
P = {P_0}\exp \left( { - \frac{{2\pi }}{\lambda }L{\mathop{\rm Im}\nolimits} \left( \chi  \right)} \right)\\
 \approx \mathop P\limits^\_  + \mathop P\limits^\_ \frac{{4\pi }}{\lambda }\frac{{{\wp _{cd}}}}{\hbar }\frac{{{N_0}\wp _{ab}^2}}{{{\varepsilon _0}\hbar {\Omega _p}}}L\cdot{\mathop{\rm Im}\nolimits} \left[ {{h_1}\left( t \right) \otimes {e^{i\omega t}} + {h_2}\left( t \right) \otimes {e^{ - i\omega t}}} \right]\\
 = \mathop P\limits^\_  + \mathop P\limits^\_ \frac{{4\pi }}{\lambda }\frac{{{\wp _{cd}}}}{\hbar }\frac{{{N_0}\wp _{ab}^2}}{{{\varepsilon _0}\hbar {\Omega _p}}}L\cdot{\mathop{\rm Im}\nolimits} \left[ {{H_1}\left( \omega  \right){e^{i\omega t}} + {H_2}\left( { - \omega } \right){e^{ - i\omega t}}} \right]\\
 = \mathop P\limits^\_  + \mathop P\limits^\_ \frac{{4\pi }}{\lambda }\frac{{{\wp _{cd}}}}{\hbar }\frac{{{N_0}\wp _{ab}^2}}{{{\varepsilon _0}\hbar {\Omega _p}}}L\cdot Amp \cdot\cos \left( {\omega t + \phi } \right).\\
 Amp = \sqrt {{{\left[ {{H_{1R}}\left( \omega  \right) - {H_{2R}}\left( { - \omega } \right)} \right]}^2} + {{\left[ {{H_{2I}}\left( { - \omega } \right) + {H_{1I}}\left( \omega  \right)} \right]}^2}}.
\end{array}
\label{eq22}
\end{equation}

In the above equations, \( H_{1R}(\omega) \), \( H_{1I}(\omega) \), \( H_{2R}(-\omega) \), and \( H_{2I}(-\omega) \) are the real and imaginary parts of \( H_1(\omega) \) and \( H_2(-\omega) \), respectively,
where

$P = P_0 \exp \left( \frac{4\pi N_0 \epsilon_{ab}^2 L}{\lambda \epsilon_b h \Omega_p} \text{Im} \left( \rho_{ba}^0 \right) \right)$, 

$\quad \phi = \tan^{-1} \frac{H_{2k}(-\omega) - H_{1k}(\omega)}{H_{2l}(-\omega) + H_{1l}(\omega)}$.

Eq. (\ref{eq22}) describes how the transmitted optical power through the atomic vapor cell varies for single-frequency electric fields of different frequencies. This variation in transmitted power reflects the bandwidth characteristics of the Rydberg receiver, which are related to both the real and imaginary parts of \( H_1(\omega) \) and \( H_2(-\omega) \). We investigate the characteristics of \( H_{1R}(\omega) \), \( H_{1I}(\omega) \), \( H_{2R}(-\omega) \), and \( H_{2I}(-\omega) \) through simulation. Experimental aspects such as the relationship between the Rydberg receiver's bandwidth and parameters will be addressed in the experimental section.
In our simulations, we focus solely on studying the characteristics of \( H_1(\omega) \) and \( H_2(\omega) \). Since the dephasing rate is difficult to estimate, we assume ideal conditions where both \( \gamma_t \) and \( \gamma_{dep} \) are set to zero. The impact of dephasing attenuation on the response will be analyzed in subsequent simulations.

\begin{figure}[t]
    \centering
    \includegraphics[width=1\linewidth]{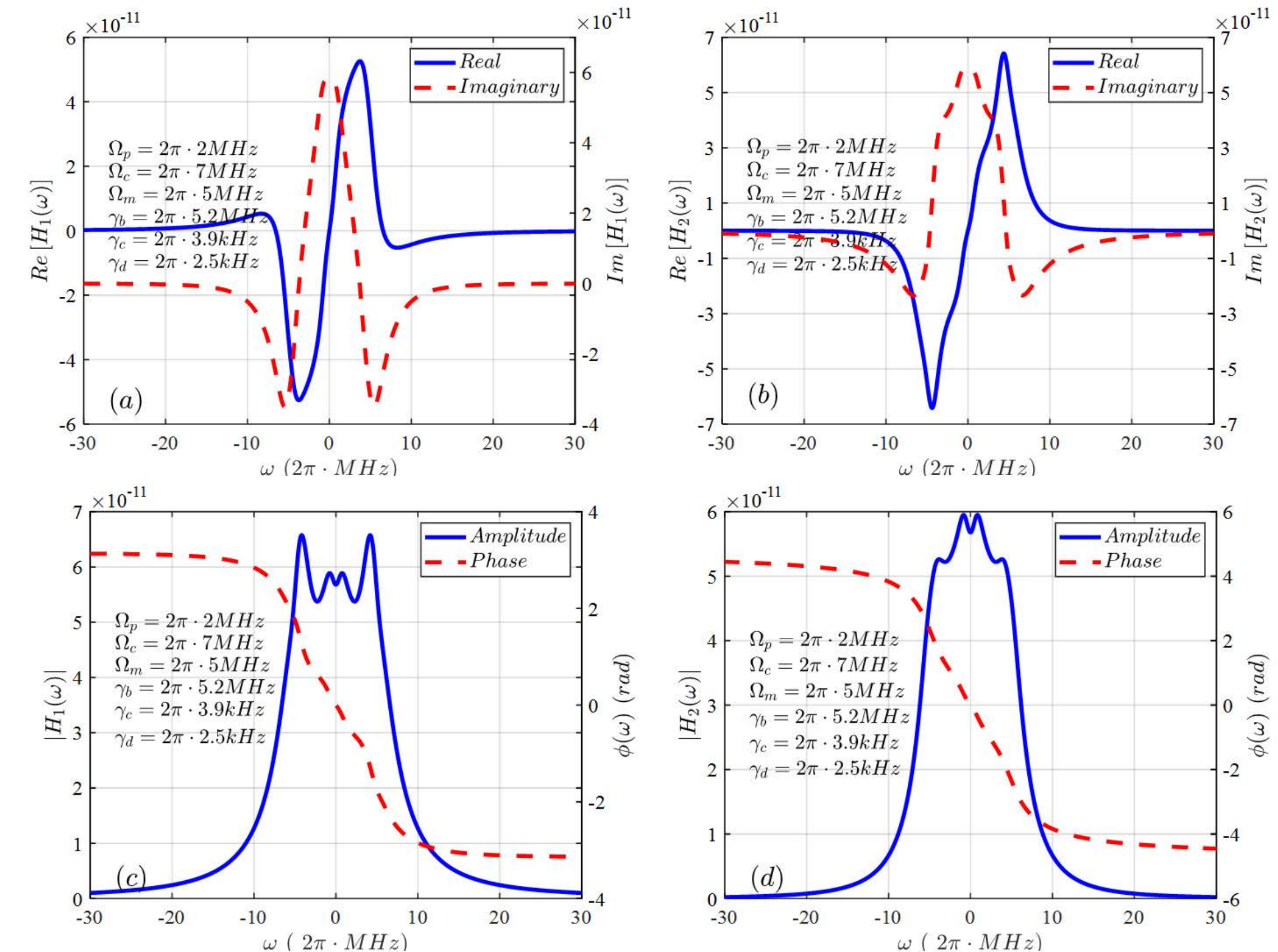}
    \caption{The real part, imaginary part, magnitude, and phase of the receiver responses \( H_1(\omega) \) and \( H_2(\omega) \).}
    \label{fig3}
\end{figure}

Fig. \ref{fig3} presents the real part, imaginary part, magnitude, and phase curves of \( H_1(\omega) \) and \( H_2(\omega) \) for a Rydberg receiver under a specific set of parameters. From Fig. \ref{fig3}(a) and Fig. \ref{fig3}(b), it can be observed that the real and imaginary parts of \( H_1(\omega) \) and \( H_2(\omega) \) exhibit symmetry. Specifically, the real part displays odd symmetry about zero frequency, while the imaginary part exhibits even symmetry about zero frequency. Fig. \ref{fig3}(c) and Fig. \ref{fig3}(d) indicate that the magnitude response of the receiver exhibits a low-pass filter performance, while the phase response within the passband shows slightly nonlinearity.

Eq. (\ref{eq21}) indicates that when the measured electric field is an amplitude-modulated signal, i.e., when $\mathcal{E}(t)$ is real, the system response of the receiver is \( H_1(\omega)+H_2(\omega) \). Therefore, we can determine the bandwidth of the Rydberg atom receiver for amplitude-modulated signals by analyzing the magnitude  of \(|H_1(\omega)+H_2(\omega)|\). However, for frequency-modulated and phase-modulated signals, the receiver bandwidth should be described by the square root term in Eq. (\ref{eq22}). Noting that the real part of the receiver response is an odd function and the imaginary part is an even function, i.e., \( H_{2R}(-\omega)=-H_{2R}(\omega)\) and \( H_{2I}(-\omega)=H_{2I}(\omega) \), we obtain the following relationship:

\begin{equation}
\begin{array}{l}
\sqrt {{{\left[ {{H_{1R}}\left( \omega  \right) - {H_{2R}}\left( { - \omega } \right)} \right]}^2} + {{\left[ {{H_{2I}}\left( { - \omega } \right) + {H_{1I}}\left( \omega  \right)} \right]}^2}} \\
 = \left| {{H_1}\left( \omega  \right) + {H_2}\left( \omega  \right)} \right|
\end{array}
\label{eq23}
\end{equation}

Eq. (\ref{eq22}) and Eq. (\ref{eq23}) demonstrate that \(|H_1(\omega)+H_2(\omega)|\) can represent the magnitude-frequency response characteristics of the Rydberg atom receiver for both amplitude-modulated signals and frequency- or phase-modulated signals.

\begin{figure*}[t]
    \centering
    \includegraphics[width=1\linewidth]{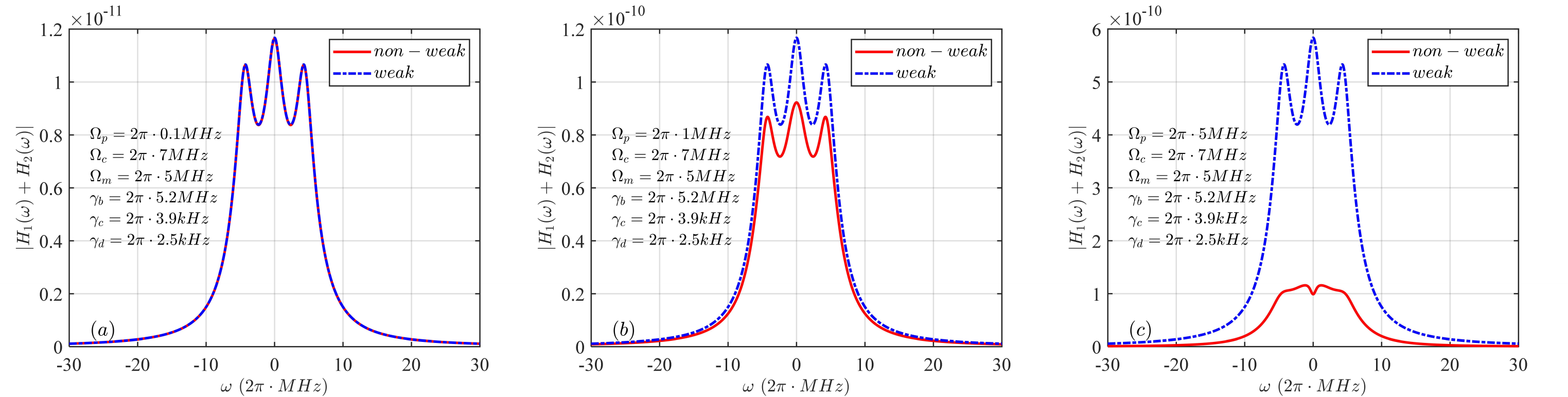}
    \caption{Comparison of the Rydberg receiver's magnitude response under non-weak-probe approximation and weak-probe approximation.}
    \label{fig4}
\end{figure*}

When the Rabi frequency of the probe light is relatively small, the response obtained using the weak-probe approximation in Eq. (\ref{eq17}) is almost identical to that obtained using Eq. (\ref{eq14}), as shown in Fig. \ref{fig4}(a). As the Rabi frequency of the probe light increases, the discrepancy between the non-weak-probe response and the weak-probe approximation becomes more significant, as illustrated in Fig. \ref{fig4}(b) and Fig. \ref{fig4}(c). Under the weak-probe approximation, the receiver response \(|H_1(\omega) + H_2(\omega)|\) is proportional to the Rabi frequency \(\Omega_p\) of the probe light. However, in the non-weak-probe solution, since \(\rho_{cb}^0\), \(\rho_{db}^0\), and \(\rho_{cc}^0 - \rho_{dd}^0\) are all non-zero, these factors cause the magnitude response \(|H_1(\omega) + H_2(\omega)|\) to decrease compared to that under the weak-probe assumption, and also affect the shape of the response.

It is desirable that changes in the external electric field to be detected induce a significant variation in the susceptibility \( \chi \) of the medium, indicating a stronger response of the Rydberg receiver to the measured field. Increasing the Rabi frequency of the probe light significantly enhances \( |H_1(\omega) + H_2(\omega)| \), as shown in Fig. \ref{fig4}(c). However, the susceptibility of the medium influenced by the measured electric field is proportional to \( |H_1(\omega) + H_2(\omega)|/\Omega_p \). As shown in Fig. \ref{fig4}, this ratio is higher when the probe light is relatively weak. Nevertheless, reducing the probe light power directly leads to a very weak average transmitted power $\overline{P}$ through the cell, which is unfavorable for photodetection of the useful signal.

\begin{figure}[h]
    \centering
    \includegraphics[width=0.9\linewidth]{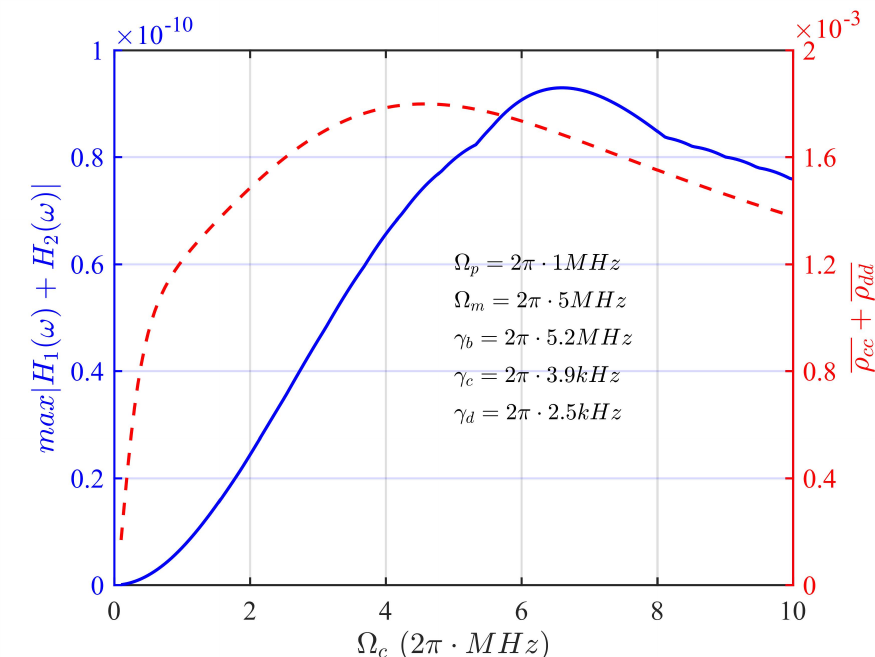}
    \caption{Effect of Varying the Coupling Laser Rabi Frequency on Receiver Gain.}
    \label{fig5}
\end{figure}

The coupling light does not directly participate in photodetection. The coupling light affects the power of the probe light by influencing the susceptibility of the mediumm, and the susceptibility of the medium is proportional to \( |H_1(\omega) + H_2(\omega)| \). Therefore, we can change the Rabi frequency of the coupling light and monitor the variation in the maximum value of \( |H_1(\omega) + H_2(\omega)| \) to illustrate the effect of \( \Omega_c \) on the receiver. The results are shown in Fig. \ref{fig5}. In Fig. \ref{fig5}, the left axis represents the maximum value of \( |H_1(\omega) + H_2(\omega)| \), and the right axis represents the average probability of atoms being in the Rydberg state when changing the Rabi frequency of the coupling light, where $\overline{\rho_{cc,dd}} = \int \rho_{cc,dd}^0 (v_z) f(v_z) dv_z $. It implies that as the Rabi frequency of the coupling light increases from zero, more atoms are excited from the intermediate state to the Rydberg state, leading to a rapid increase in both the population of the Rydberg level and the receiver's response. As the Rabi frequency of the coupling light continues to increase, both the Rydberg state population and the magnitude response of the receiver exhibit a peak. The peak of the Rydberg state population occurs earlier than that of the receiver's magnitude response. Under the simulation parameters in the figure, the peak of the receiver occurs at a coupling light Rabi frequency of approximately 6.5 MHz. Due to the very small transition dipole moment between the intermediate state and the Rydberg state, a coupling light with higher power should be selected.

\begin{figure}[h]
    \centering
    \includegraphics[width=1\linewidth]{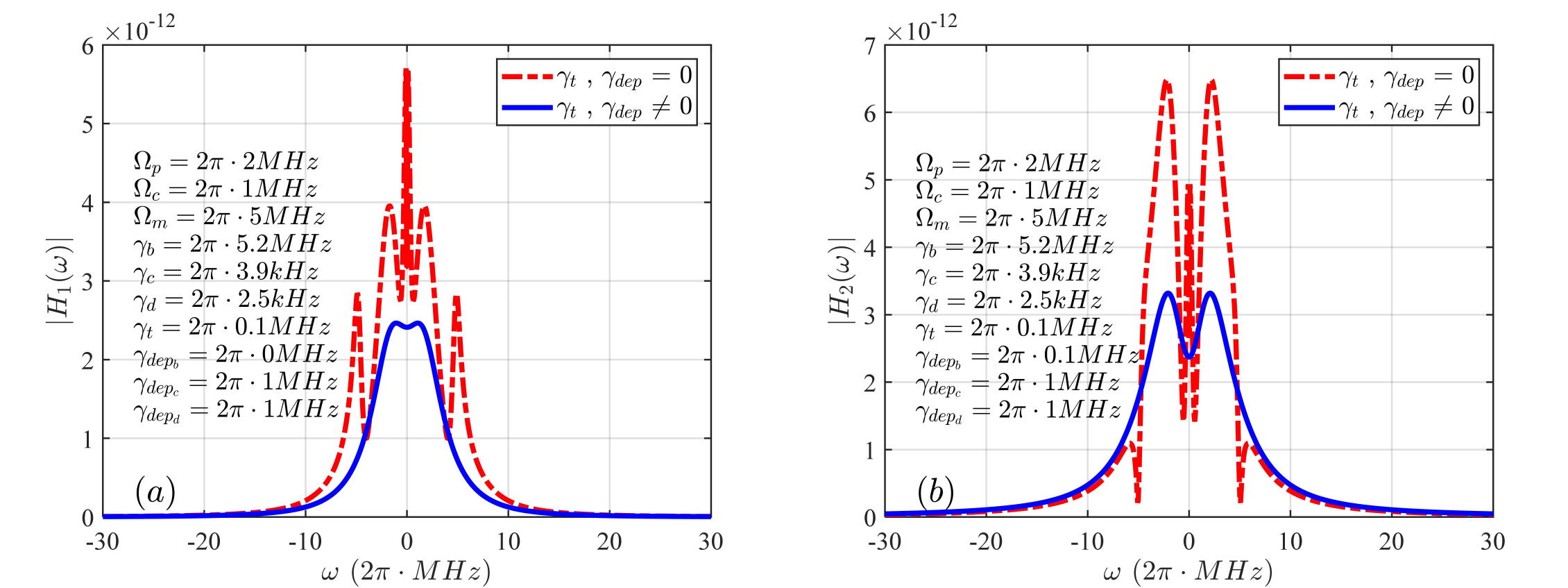}
    \caption{Effect of Dephasing and Transit-time Decay on the Response.}
    \label{fig6}
\end{figure}

In the simulations above, all calculations were performed under ideal conditions, meaning that transit-time decay \( \gamma_t \) and the dephasing decay rates of each energy level—\( \gamma_{dep\_b} \), \( \gamma_{dep\_c} \), and \( \gamma_{dep\_d} \)—were not considered. However, in practice, dephasing decay cannot be neglected and has a significant impact on the system's response. We simulated this scenario, and the results are shown in Fig. \ref{fig6}. The red line in the figure represents the simulation without transit-time decay and dephasing decay. It can be observed that, under the selected parameters, both \( |H_1(\omega)| \) and \( |H_2(\omega)| \) exhibit severe oscillations, making the response of the atomic receiver appear chaotic.

Transit-time decay is primarily influenced by the laser beam diameter, while dephasing decay is mainly caused by collisions between atoms. State \( |b\rangle \) is generally chosen as the first excited state of alkali metal atoms, where the binding energy from the atomic nucleus is relatively large, and collisions between atoms are less likely to alter its state. Therefore, in the simulation, we set \( \gamma_{dep\_b} = 0 \). States \( |c\rangle \) and \( |d\rangle \) are Rydberg states, where atoms have extremely large diameters and the binding energy from the atomic nucleus is very small. Collisions due to atomic thermal motion are sufficient to alter the atomic states. In the simulation, we selected \( \gamma_{dep\_c,d} = 2\pi \times 1\ \text{MHz} \).

When transit-time decay and dephasing decay are included, the magnitude of the response changes significantly. The originally severe oscillations within the passband become smooth, and the atomic receiver's response has a low-pass filter-like characteristic as shown in Fig. \ref{fig6}. Moreover, when considering the transit decay and dephasing decay, the magnitude of the atomic receiver's response decreases noticeably. Both the smoothing effect within the passband and the reduction in response magnitude are attributed to the dephasing decay disrupting the coherence of the four-level atom and reducing its sensitivity.

\section{Intrinsic Noise Performance of $\rho_{ba}$ }

In the previous section, we derived in detail the analytical expression for the dynamic solution of the density matrix element \(\rho_{ba}\) in a four-level heterodyne Rydberg receiver. This dynamic solution is related to the static solution and the time-varying term introduced by the electric field to be received. Assuming the electric field to be received is a single-frequency field with a detuning of \(\omega\) from the two Rydberg levels and an amplitude of \(E_s\), the expression for \(\rho_{ba}\) of an atom with a motion velocity of \(v_z\) is given by:

\begin{equation}
\begin{aligned}
    \rho_{ba}\left(t,v_z\right)&=\rho_{ba}^0+\frac{\wp_{cd}E_s}{\hbar}\left(H_{1v}\left(\omega\right)e^{i\omega t}+H_{2v}\left(-\omega\right)e^{-i\omega t}\right).
\end{aligned}
\label{eq24}
\end{equation}

In this section, we analyze the noise performance of \(\rho_{ba}\). The expression for \(\rho_{ba}\) consists of three terms, where \(\rho_{ba}^{(0)}\) is the density matrix element in the absence of the microwave electric field to be received, and the other two terms are time-dependent components introduced by the received electric field. When the Rabi frequency of the received electric field is much smaller than that of the local oscillator microwave field, the absolute value of \(\rho_{ba}^{(0)}\) is significantly larger than those of the other two terms. Therefore, when discussing the noise of \(\rho_{ba}\), the noise performance of \(\rho_{ba}^{(0)}\) dominate. The analytical expression for \(\rho_{ba}^{(0)}\) in Eq. (\ref{eq14}) indicates that \(\rho_{ba}^{(0)}\) depends on the population distribution probabilities of each energy level, the Rabi frequencies of the probe light, coupling light, and local oscillator microwave field, as well as the decay rates of each energy level. When we disregard the quantum noise of the probe light, coupling light, and local oscillator microwave field—that is, when the four-level atom is excited by ideal classical light sources—the noise in \(\rho_{ba}^{(0)}\) is solely related to the population probabilities of each energy level. This noise reflects the intrinsic noise of the four-level atom itself.

In the four-level atom, the population probabilities of particles in each energy level are determined by the Rabi frequencies of the probe light, coupling light, and local oscillator microwave field, as well as the decay rates of each energy level. The distribution of particles among the energy levels follows a multinomial distribution. Assuming \(n_i\) is the particle density in each energy level:

\begin{equation}
\begin{array}{l}
P\left\{ {{Y_a} = {n_a},{Y_b} = {n_b},{Y_c} = {n_c},{Y_d} = {n_d}} \right\}\\
\begin{array}{*{20}{c}}
\end{array} = \frac{{N!}}{{{n_a}!{n_b}!{n_c}!{n_d}!}}{\left( {\rho _{aa}^0} \right)^{{n_a}}}{\left( {\rho _{bb}^0} \right)^{{n_b}}}{\left( {\rho _{cc}^0} \right)^{{n_c}}}{\left( {\rho _{dd}^0} \right)^{{n_d}}}\begin{array}{*{20}{c}},
{}&{}
\end{array}
\end{array}
\label{eq25}
\end{equation}
where $n_a+n_b+n_c+n_d = N$ and $\rho_{aa}^0+\rho_{bb}^0+\rho_{cc}^0+\rho_{dd}^0=1$.

For a multinomial distribution, the expectation of the particle density in a given energy level is \( E(Y_i) \), the variance of the particle concentration is \( \sigma^2(Y_i) \), and the expectation of the product of particle densities in two energy levels is \( E(Y_i Y_j) \).

\begin{equation}
\begin{array}{l}
E\left( {{Y_i}} \right) = N\rho _{ii}^0, \\
{\sigma ^2}\left( {{Y_i}} \right) = E\left( {Y_i^2} \right) - {E^2}\left( {{Y_i}} \right) = N\rho _{ii}^0\left( {1 - \rho _{ii}^0} \right), \\
E\left( {{Y_i}{Y_j}} \right) = N\left( {N - 1} \right)\rho _{ii}^0\rho _{jj}^0. 
\end{array}
\label{eq26}
\end{equation}

Eq. (\ref{eq14}) indicates that \( \rho_{ba}^0 \) is related to the population difference between energy levels. The population difference leads to a concentration difference of particles between two energy levels. Since the particle concentration in each energy level is a random variable, the difference in particle concentrations between two energy levels is also random. Using Eq. (\ref{eq26}), the variance of the difference in particle concentrations between two energy levels can be readily derived.

\begin{equation}
{\sigma ^2}\left( {{Y_i} - {Y_j}} \right) = N\left( {\left( {\rho _{ii}^0 + \rho _{jj}^0} \right) - {{\left( {\rho _{ii}^0 - \rho _{jj}^0} \right)}^2}} \right).
\label{eq27}
\end{equation}

Eq. (\ref{eq27}) indicates that under the multinomial distribution condition, the variance of the particle concentration difference between adjacent energy levels depends only on the distribution probabilities of particles in these two energy levels and is independent of the distribution probabilities in other energy levels. We need to depict the noise performance of the density matrix element \(\rho_{ba}^0\), i.e., to describe the standard deviation of \(\rho_{ba}^0\). The expression for the density matrix element \(\rho_{ba}^0\) shows that the only random variables that can introduce stochasticity into \(\rho_{ba}^0\) are the population probability differences of adjacent energy levels, while all other factors are deterministic constants. We can multiply both the numerator and denominator of \(\rho_{ba}^0\) by the particle concentration, transforming the population probability difference between two adjacent energy levels into the difference in particle population concentrations between these two levels. Furthermore, \(\rho_{ba}^0\) can be rewritten as a weighted summation of the differences in particle concentrations between adjacent energy levels. That is, Eq. (\ref{eq14b}) can be reformulated as:

\begin{equation}
\rho _{ba}^0\left( {{v_z}} \right) = {{\rm{{\cal A}}}_1}\frac{{{Y_{ab}}}}{{N\left( {{v_z}} \right)}} + {{\rm{{\cal A}}}_2}\frac{{{Y_{bc}}}}{{N\left( {{v_z}} \right)}} + {{\rm{{\cal A}}}_3}\frac{{{Y_{cd}}}}{{N\left( {{v_z}} \right)}}.
\label{eq28}
\end{equation}

where:
$\mathcal {A}_1 = \frac{{{\Omega _p}}}{{\left( {i\gamma _{ba}^{\prime} - 2{k_p}{v_z} - \frac{{\Omega _c^2}}{{E\left( {{k_p} - {k_c}} \right)}}} \right)}}$,\\

$\mathcal{A}_2 =  - \frac{{{\Omega _p}\Omega _c^2\left( {1 + \left( {\frac{{\Omega _L^2}}{{A\left( { - {k_c}} \right)D\left( {{k_p} - {k_c}} \right)}}} \right)\left( {\frac{{\Omega _p^2}}{{A\left( { - {k_c}} \right)B\left( { - {k_c}} \right)}} + 1} \right)} \right)}}{{B\left( { - {k_c}} \right)E\left( {{k_p} - {k_c}} \right)\left( {i\gamma _{ba}^{\prime} - 2{k_p}{v_z} - \frac{{\Omega _c^2}}{{E\left( {{k_p} - {k_c}} \right)}}} \right)}}$,\\

$\mathcal{A}_3 = {\Omega _p}\Omega _c^2\left( {\frac{\begin{array}{l}
\left( {1 + \frac{{\Omega _L^2}}{{A\left( { - {k_c}} \right)B\left( { - {k_c}} \right)}}} \right)\cdot\left( {1 + \frac{{\Omega _p^2}}{{A\left( { - {k_c}} \right)B\left( { - {k_c}} \right)}}} \right)\\
\cdot\left( {\frac{{\Omega _L^2}}{{A\left( { - {k_c}} \right)D\left( {{k_p} - {k_c}} \right)}}} \right) + \left( {\frac{{\Omega _L^2}}{{A\left( { - {k_c}} \right)B\left( { - {k_c}} \right)}}} \right)
\end{array}}{{i\gamma _{dc}^{\prime}E\left( {{k_p} - {k_c}} \right)\left( {i\gamma _{ba}^{\prime} - 2{k_p}{v_z} - \frac{{\Omega _c^2}}{{E\left( {{k_p} - {k_c}} \right)}}} \right)}}} \right).$

In the above equation, $Y_{ij}$ represents the particle concentration difference between the $i$ and $j$ energy levels, satisfying $Y_{ij} = N(v_z)(\rho_{ii}^0 - \rho_{jj}^0)$, The particle concentration difference depends on the atomic velocity and is a function of $v_z$. Since the cross terms in the variance calculation are relatively small, the variance of the density matrix element $\rho_{ba}^0(v_z)$ can be approximately expressed as:

\begin{equation}
\begin{array}{l}
{\sigma ^2}\left( {\rho _{ba}^0\left( {{v_z}} \right)} \right) \approx \frac{{{{\left| {{{\rm{{\cal A}}}_1}} \right|}^2}}}{{{N^2}\left( {{v_z}} \right)}}{\sigma ^2}\left( {{Y_{ab}}} \right) + \frac{{{{\left| {{{\rm{{\cal A}}}_2}} \right|}^2}}}{{{N^2}\left( {{v_z}} \right)}}{\sigma ^2}\left( {{Y_{bc}}} \right)\\
\begin{array}{*{20}{c}}
{\begin{array}{*{20}{c}}
{\begin{array}{*{20}{c}}
{}&{}
\end{array}}&{}&{}
\end{array}}&{}&{}&{}
\end{array} + \frac{{{{\left| {{{\rm{{\cal A}}}_3}} \right|}^2}}}{{{N^2}\left( {{v_z}} \right)}}{\sigma ^2}\left( {{Y_{cd}}} \right)
\end{array}.
\label{eq29}
\end{equation}

By taking the weighted integral over the velocity distribution, the average standard variance of the density matrix element$\rho_{ba}^0$ can be obtained.

\begin{equation}
\sigma \left( {\overline {\rho _{ba}^0} } \right) = \int {\sigma \left( {\rho _{ba}^0\left( {{v_z}} \right)} \right)f\left( {{v_z}} \right)d{v_z}}.
\label{eq30}
\end{equation}

So far, we have derived the expression for the standard deviation of the density matrix element \( \rho_{ba}^0 \). How does this standard deviation affect the sensitivity of the atomic receiver? From Eq. (\ref{eq14a}), it is known that the density matrix element \( \rho_{ba} \) depends not only on \( \rho_{ba}^0 \), but also on the receiver responses \( H_1(\omega) \) and \( H_2(\omega) \). When receiving a single-frequency electric field with a detuning of \( \omega \) from the two Rydberg levels and an amplitude of \( E_s \), the following relationship holds after considering the weighted integral over the atomic velocities:

\begin{equation}
\overline {{\rho _{ba}}} \left( t \right) \approx \overline {\rho _{ba}^0}  + \frac{{{\wp _{cd}}{E_s}}}{\hbar }\left({H_1}\left( \omega  \right){e^{i\omega t}} + {H_2}\left( { - \omega } \right){e^{ - i\omega t}}\right).
\label{eq31}
\end{equation}

In the above equation:  $\overline {\rho_{ba}}(t) = \int \rho_{ba}(t,v_z)f(v_z)dv_z$ . As mentioned earlier, \(|H_1(\omega) + H_2(\omega)|\) represents the receiver's gain for a signal at frequency \(\omega\). Eq. (\ref{eq31}) consists of two parts: the first part is a random variable with a mean of $\overline{\rho_{ba}^0}$ and a standard deviation of \(\sigma(\overline{\rho_{ba}^0})\); the second part is a single-frequency signal related to the Rabi frequency of the measured signal. For the receiver to detect the single-frequency signal, the effective value of the single-frequency signal must be greater than the standard deviation of the noise at that frequency point. Assuming the receiver has a flat magnitude-frequency response with bandwidth \(B\), the standard deviation of the noise per unit bandwidth is \(\sigma(\overline{\rho_{ba}^0})/\sqrt{B}\). Therefore, the minimum detectable field strength (RMS value) of the receiver is:

\begin{equation}
{E_{\min }} = \frac{{\hbar \sigma \left( {\overline {\rho _{ba}^0} } \right)}}{{\sqrt 2 {\wp _{cd}}\left| {{H_1}\left( 0 \right) + {H_2}\left( 0 \right)} \right|\sqrt B }}.
\label{eq32}
\end{equation}

The receiver sensitivity described by Eq. (\ref{eq32}) is derived without considering detector noise, laser intensity noise and laser phase noise. This sensitivity is determined by the random distribution of particles across energy levels, reflecting the intrinsic noise of the four-level. Such noise cannot be eliminated and is referred to as the intrinsic noise of the Rydberg receiver. Therefore, the sensitivity described by Eq. (\ref{eq32}) represents the fundamental sensitivity limit of the Rydberg receiver.

\section{Experimental}

\begin{figure}[h]
    \centering
    \includegraphics[width=1\linewidth]{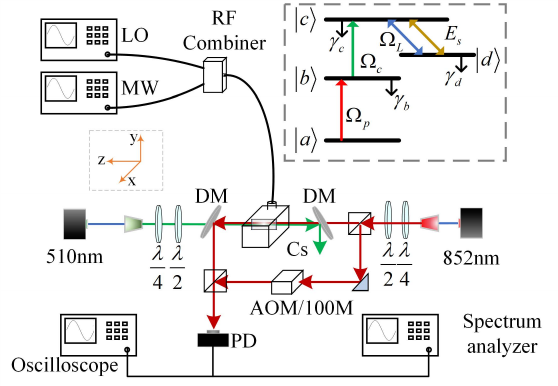}
    \caption{Schematic diagram of the experimental setup and energy level structure of the heterodyne Rydberg receiver. LO represents the local oscillator microwave field, and MW denotes the weak target microwave field, which are coupled via a radio-frequency combiner before entering the waveguide. A combination of a half-wave plate and a quarter-wave plate is used to adjust the polarization of the light field. A dichroic mirror (DM) reflects light of specific wavelengths, and a photodetector (PD) captures the transmitted signal of the probe light. The inset in the upper right corner illustrates the energy level structure of the four-level heterodyne Rydberg atom receiver.}
    \label{fig7}
\end{figure}

In this section, we conduct practical measurements of the response of the four-level heterodyne Rydberg receiver. The experimental setup is shown in Fig. \ref{fig7}. We use cesium atoms in our experiment. The frequencies of the 852 nm laser and the 510 nm laser are locked to the atomic resonance frequency using a dual-wavelength ultra-stabilization system. The 852 nm probe light is generated by a Moglabs cat's eye laser. The 510 nm coupling light is obtained by frequency doubling the output of a Moglabs 1020 nm laser using a amplifier and frequency doubler (FL-SF-509-1-CW) from Shanghai Precilasers. The ultra-stabilization system is a VH6020-4 dual-wavelength system from SLS.

The 852 nm probe light and the 510 nm coupling light are incident on the cesium vapor cell in a counter-propagating configuration (the cesium cell used in the experiment is a 1.5 cm cube) to excite the cesium atoms to the Rydberg state. The $e^{-2}$ beam diameter of the probe light illuminating the cesium cell is about 1.18 mm, and the $e^{-2}$ beam diameter of the coupling light is about 0.84 mm.

In the experiment, the local oscillator microwave field and the weak microwave field to be measured (provided by two Rohde Schwarz  SMB100A signal generators, respectively) are combined by a power combiner (ZN2PD-9G-S+ from Mini-Circuits) and then fed into a rectangular waveguide (WR137) containing the atomic vapor cell. The main reason for placing the cesium vapor cell inside the waveguide is that the polarization of the electromagnetic field propagating in a rectangular waveguide is predominantly vertical. This allows us to conveniently control the polarization directions of the probe and coupling lights to be parallel to the polarization of the microwave field. Placing the cell inside the rectangular waveguide effectively shields it from the influence of various stray electromagnetic waves in the experimental environment that could affect the polarization state of the microwave field under test.

The probe light transmitted through the cesium cell is combined with a local oscillator beam. This local oscillator beam is the 852 nm probe light frequency-shifted by an AOM (MT110-B50-A1.5-IR from AA Opto-Electronique, France). The combined light is then detected by an APD detector (APD410A from Thorlabs). The output of the photodetector is connected to a spectrum analyzer (RS FSWR26, 20 Hz–26.5 GHz) for measurement. The clocks of the signal generators, oscilloscope, and spectrum analyzer are synchronized during the measurements.

In the experiment, we measured the atomic receiver response under different parameters by varying the probe laser power and coupling laser power, thereby changing the corresponding Rabi frequencies, and compared the results with theoretical simulations. The Rydberg states we selected are the $47D_{5/2}$ state and the $48P_{3/2}$ state. A key difference from the previous simulations is that the Rabi frequency can no longer be treated as constant throughout the light propagation. As the probe light travels through the atomic vapor cell, its power decreases with increasing propagation distance. Consequently, the Rabi frequency of the probe light is not constant but gradually decreases along the path. However, since the transition dipole moment between the $48P_{3/2}$ state and the $47D_{5/2}$ state is very small (approximately $0.0187 e a_0$), thus the absorption of the coupling light by the medium is very weak, so the Rabi frequency of the coupling light can be considered as constant.

In the simulation, we divided the length of the atomic vapor cell into 30 segments along the propagation direction. Within each segment, the Rabi frequency of the probe light was assumed to be constant. Furthermore, because the laser beam has a Gaussian profile, we calculated the average Rabi frequency within the Gaussian beam from the laser power using:

\begin{equation}
\overline \Omega{p,c}   = \frac{{8\wp_{ab,bc} \sqrt {P_{p,c}} }}{{3\hbar \sqrt {c{\varepsilon _0}\pi w_0^2} }},
\label{eq33}
\end{equation}
where $w_0$ is the $e^-2$ diameter of the probe beam or coupling beam.

We place the cesium cell inside the rectangular waveguide. The mode propagating in the rectangular waveguide is TE10, and the amplitude of the electric field inside the waveguide has a fixed analytical expression, so we can calculate the electric field amplitude by:

\begin{equation}
P = E_0^2\frac{{ab}}{4}\sqrt {\frac{{{\varepsilon _0}}}{{{\mu _0}}}} \sqrt {1 - {{\left( {\frac{c}{{2af}}} \right)}^2}}. 
\label{eq34}
\end{equation}

In the above equation, \( a \) and \( b \) represent the lengths of the long and short sides of the rectangular waveguide, respectively. To ensure the electric field within the waveguide remains in a traveling-wave state, the output port of the waveguide must be terminated with a 50-ohm load. In our system, there is a total loss of 10.5 dB from the set power of the signal source to the traveling-wave power transmitted in the waveguide. This includes the 6.5 dB insertion loss of the ZN2PD-9G-S+ power combiner, cable connection losses, and the losses introduced by placing the atomic vapor cell inside the waveguide.

Although in our theoretical derivation we assumed that the frequencies of the probe light, coupling light, and microwave field are exactly resonant with their respective energy levels without any detuning, in actual experiments, we first use an ultrastable cavity to lock the probe light frequency at 351.72196 THz (using a HighFinesse WS7-60 wavemeter with an accuracy of 60 MHz). Due to the limited accuracy of the wavemeter and factors such as wavemeter calibration, we cannot guarantee that the actual frequency of the probe light exactly matches the transition frequency between the \(6S_{1/2}(F=4)\) and \(6P_{3/2}(F=5)\) states of the cesium atom, resulting in a certain detuning \(\Delta_p\). We then scan the frequency of the coupling laser and observe the electromagnetically induced transparency (EIT) signal. Using the ultrastable cavity, we lock the coupling laser frequency to the peak of the EIT spectrum. At this point, the coupling laser frequency also has a certain detuning \(\Delta_c\), which is related to the probe detuning by: \(\Delta_c = -\left(\lambda_p/\lambda_c\right)\Delta_p\).

When we apply the microwave field and scan the coupling laser frequency again, EIT-AT splitting is observed near the transition frequency between the two Rydberg states. We adjust the microwave frequency until the heights of the left and right peaks of the EIT-AT splitting are equal. Due to the detuning of the probe light, the peak of the EIT signal does not occur exactly at the resonant frequency between the \(6P_{3/2}\) and \(47D_{5/2}\) states with detuning of  \(\Delta_c\). Consequently, the microwave frequency also exhibits a detuning from the resonant frequency between the Rydberg states. Although the probe light, coupling light, and microwave local oscillator field all have detunings, the shape of the EIT-AT spectrum remains essentially the same as in the detuning-free case. We refer to this condition as equivalent resonance.

As mentioned earlier, the response of the Rydberg atom receiver is closely related to the dephasing rate, which is primarily caused by collisions between atoms. These include collisions between ground-state atoms and Rydberg atoms, collisions between the intermediate state \(6P_{3/2}\) and the Rydberg states \(48P_{3/2}\) and \(47D_{5/2}\), as well as collisions between Rydberg atoms themselves. The collision cross-sections for these processes vary with some dominated by the C3 coefficient and others by the C6 coefficient. Furthermore, the average collision number depends on the atomic concentration in each state. In the four-level system, the atomic concentration in each state is related to the Rabi frequencies of the probe light, coupling light, and local oscillator field. Precise modeling of all collision mechanisms is impractical.

We consider that the binding energy of the outermost electron in the \(6P_{3/2}\) state is relatively strong, and its atomic radius is small, so collisions are unlikely to alter its state. Therefore, we set \(\gamma_{{dep\_b}} = 0\). In contrast, the binding energy of the outermost electron in Rydberg states is weak, and the atomic radius is very large. Collisions can easily change the state of a Rydberg atom, making them highly susceptible to inelastic collisions. Thus, we primarily consider the dephasing rate of the Rydberg states. For simplicity, we assume \(\gamma_{{dep\_c}} = \gamma_{{dep\_d}}\). In the experiment, we selected \(\gamma_{{dep\_c}} = \gamma_{{dep\_d}} = 2\pi \times 3\ \text{MHz}\), as the simulation results show good agreement with experimental measurements under this parameter. The transit-time decay rate is \(\gamma_t = 2\left(\frac{\sigma(v_t)}{\omega_0}\right)\sqrt{2\ln(2)} = 2\pi \times 0.09\ \text{MHz}\).

\begin{figure*}[t]
    \centering
    \includegraphics[width=1\linewidth]{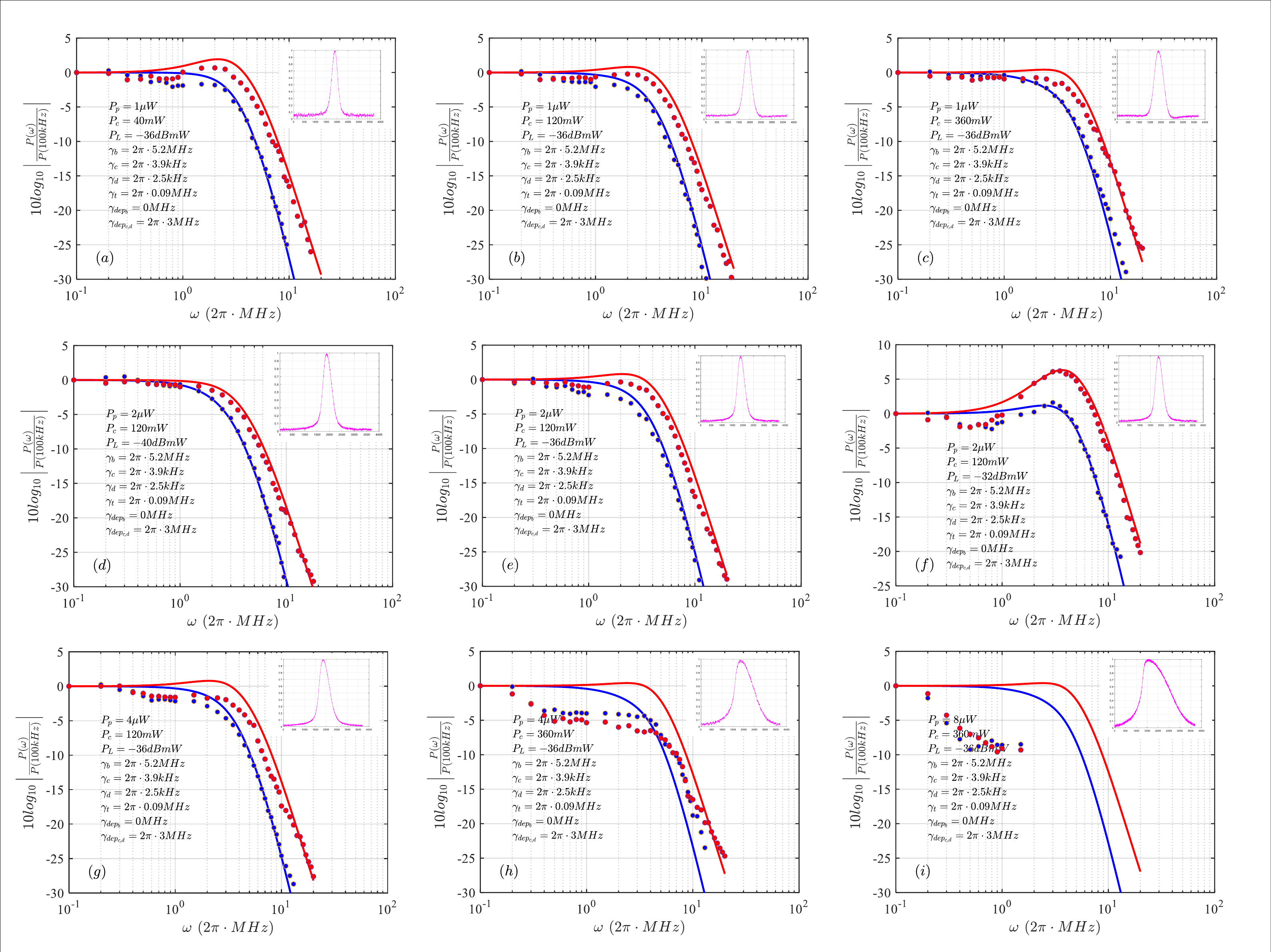}
    \caption{Measured and Simulated Results of Left and Right Sideband Power..}
    \label{fig8}
\end{figure*}

Fig. \ref{fig8} shows a comparison between the measured sideband power and simulation results under different parameters. In the figure, the blue data represents the right sideband, and the red data represents the left sideband. The frequencies of the probe laser, coupling laser, and local oscillator field remain fixed. The horizontal axis represents the frequency difference between the measured signal electric field and the local oscillator field. The curve in the upper right corner of each plot is the EIT signal observed by scanning the coupling laser frequency without applying the microwave local oscillator field under the corresponding parameter set.

It impiles that when the EIT transparency peak is not severely distorted, the simulation results agree well with the experimental measurements, as shown in Fig. \ref{fig8}(a) to Fig. \ref{fig8}(f). However, when both the probe and coupling light powers are relatively high, significant distortion occurs in the EIT transparency peak, leading to noticeable discrepancies between the simulation and experimental data, as shown in Fig. \ref{fig8}(g) to Fig. \ref{fig8}(i). (When the probe laser power was $8 \mu W$, the optical power transmitted through the cesium vapor cell had exceeded the saturation power of the APD. Consequently, an additional 10 dB optical attenuator was inserted in front of the APD.) This discrepancy is likely due to complex excitation blockade effects and many-body effects that arise when both the probe and the  coupling Rabi frequencies are large \cite{2022Enhanced}\cite{PhysRevX.10.021023}. These nonlinear effects can cause the behavior of the four-level heterodyne receiver to deviate substantially from our system model, thereby resulting in the observed differences between the measured data and simulation results.

\begin{figure*}[t]
    \centering
    \includegraphics[width=1\linewidth]{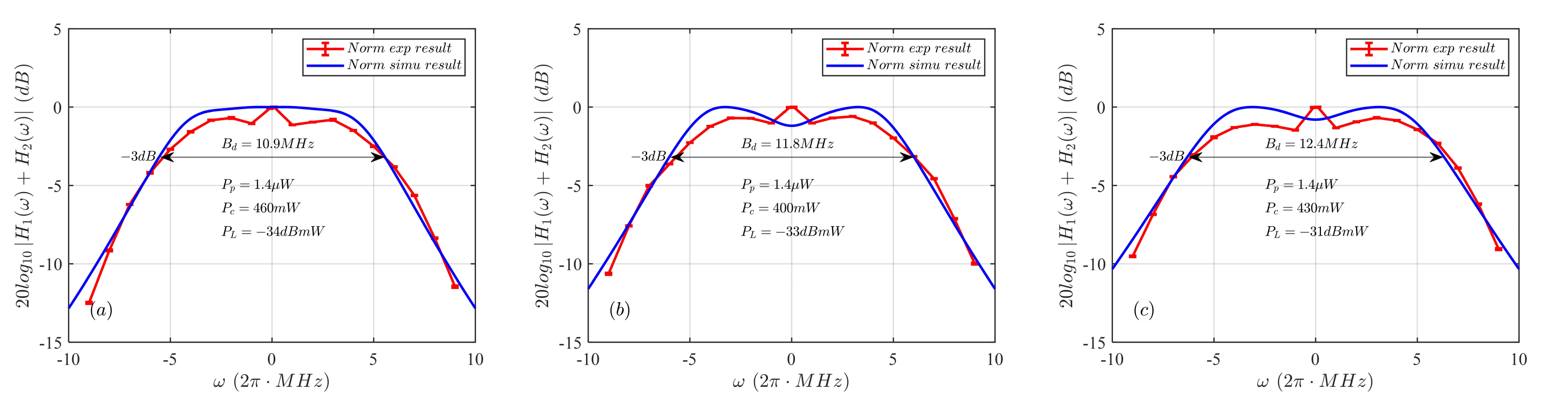}
    \caption{Comparison of Measured and Simulated System Responses.}
    \label{fig9}
\end{figure*}

We also measured the maximum achievable bandwidth of the Rydberg atomic receiver. The WR137 rectangular waveguide used in our experiment has an operational frequency range from 5.85 GHz to 8.20 GHz. Three pairs of Rydberg state atomic transitions fall within this frequency range: \(49D_{5/2} \rightarrow 50P_{3/2}\), \(47D_{5/2} \rightarrow 48P_{3/2}\), and \(45D_{5/2} \rightarrow 46P_{3/2}\), with corresponding transition frequencies of 6.0894 GHz, 6.9472 GHz, and 7.9790 GHz, respectively. The measurement results are shown in Fig. \ref{fig9}. It can be seen from the figure that the double-sideband bandwidth of the heterodyne four-level Rydberg atomic receiver can exceed 10 MHz (The definition of bandwidth as the frequency point at which the normalized response curve drops to -3 dB). Furthermore, the theoretical calculations based on the dynamic solution show good agreement with the actual measurement results.

\begin{figure}[t]
    \centering
    \includegraphics[width=1\linewidth]{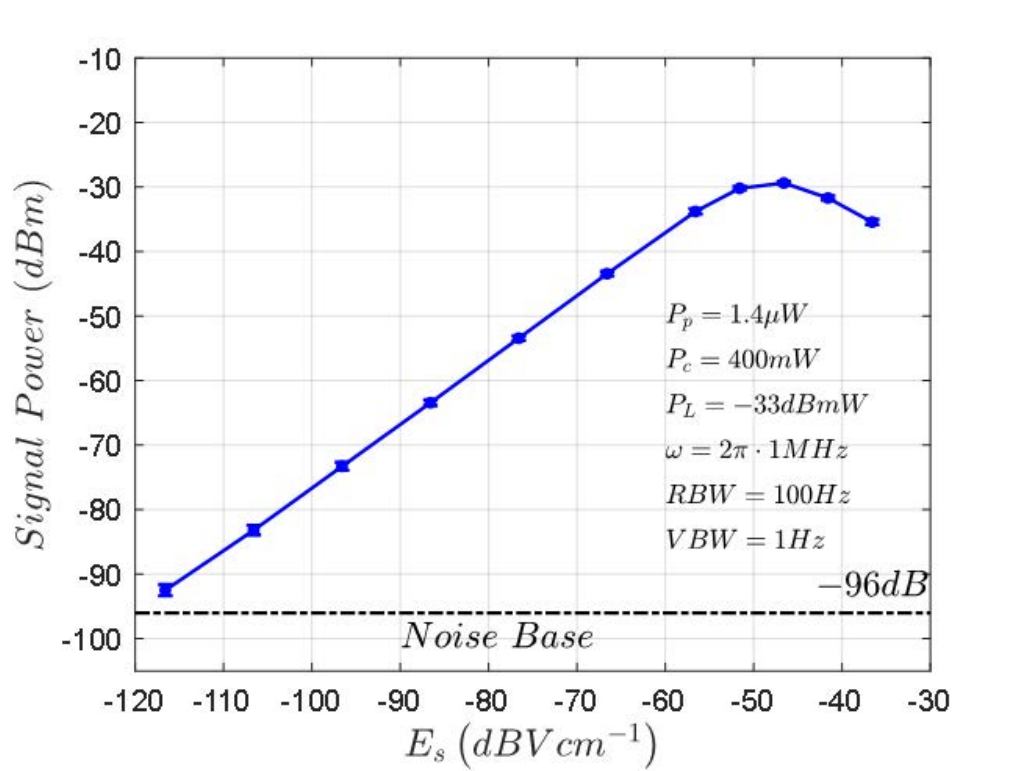}
    \caption{Dynamic Range and Sensitivity Measurements of the Rydberg Receiver.}
    \label{fig10}
\end{figure}

We tested the dynamic range of the Rydberg atomic receiver. For this test, the selected Rydberg states were \(47D_{5/2}\) and \(48P_{3/2}\). The frequency difference between the local oscillator field and the signal field was set to \(1\ \text{MHz}\), and a \(1.5\ \text{cm}\) cubic atomic vapor cell was used. The test results are shown in Fig. \ref{fig10}. The horizontal axis represents the root mean square (RMS) value of the signal electric field propagating in the waveguide, and the vertical axis represents the power read from the spectrum analyzer.

During the test, the spectrum analyzer's resolution bandwidth (RBW) was set to \(100\ \text{Hz}\) and the video bandwidth (VBW) was set to to \(1\ \text{Hz}\). Under these settings, the noise floor of the spectrum analyzer with an RBW of \(100\ \text{Hz}\) was \(-96\ \text{dBm}\). When the signal electric field strength was \(-116.5\ \text{dBV}\cdot\text{cm}^{-1}\), the corresponding signal power measured was \(-92.48\ \text{dBm}\). From this, it can be deduced that, using the \(1.5\ \text{cm}\) atomic vapor cell, the sensitivity of our four-level superheterodyne receiver is approximately \(10\ \text{nV}\cdot\text{cm}^{-1}\cdot\text{Hz}^{-1/2}\). Furthermore, the linear dynamic range of the receiver under these conditions is about \(85\ \text{dB}\).

Finally, we conducted a reception experiment using a linear frequency-modulated (chirp) signal with a bandwidth of 10 MHz. The selected Rydberg states were \(47D_{5/2}\) and \(48P_{3/2}\), and the receiver parameters were consistent with those used in the dynamic range measurement. The signal to be received was a chirp signal with a carrier frequency of \(6.9472\ \text{GHz}\) (identical to the frequency of the local oscillator field), a bandwidth of 10 MHz, a pulse width of \(2\ \text{ms}\), and a chirp rate of \(5 \times 10^9\ \text{Hz/s}\).

\begin{figure*}[t]
    \centering
    \includegraphics[width=1\linewidth]{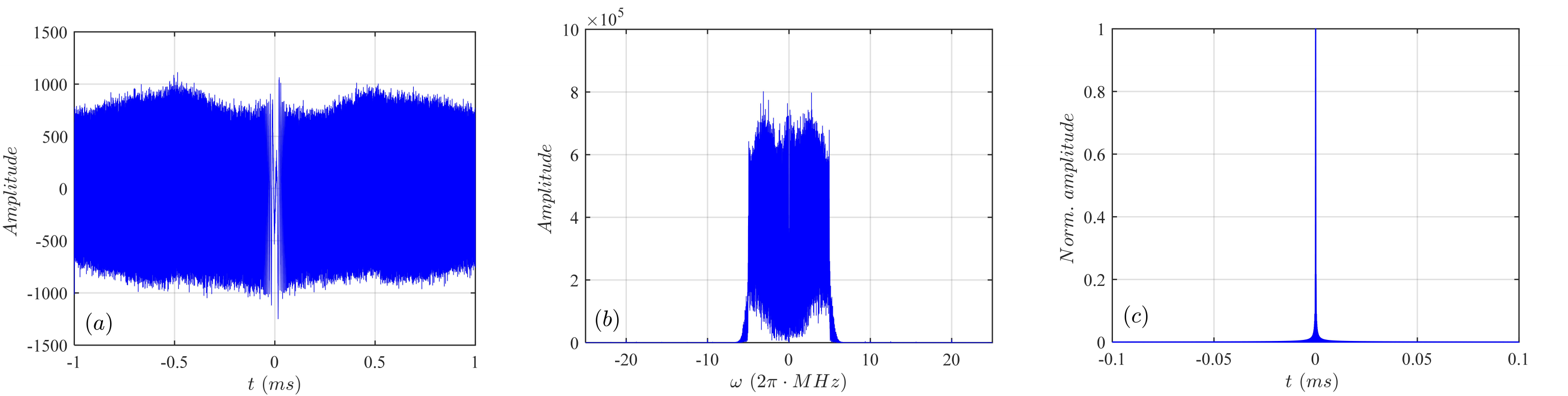}
    \caption{Reception Experiment of LFM Signals Using a Rydberg Atomic Receiver.}
    \captionsetup{justification=centering}
    \label{fig11}
\end{figure*}

The electrical signal after heterodyne detection underwent amplification, quadrature demodulation, and data acquisition. The data sampling frequency was \(50\ \text{MHz}\). The results are shown in Fig. \ref{fig11}. Fig. \ref{fig11}(a) displays the recorded chirp signal captured by the data acquisition system. Fig. \ref{fig11}(b) shows the spectrum of the received data, and Figure \ref{fig11}(c) presents the pulse compression result of the received chirp signal. It implies that the Rydberg receiver is capable of receiving modulated signals with a bandwidth of 10 MHz.

\section{Conclusion and Extension}
This paper presents a systematic study of the frequency response characteristics of the four-level heterodyne Rydberg atom receiver, proposing a method for solving and analyzing the dynamic solution of the density matrix elements. This approach successfully obtains the frequency response function of the receiver to the microwave electric field under thermal atomic conditions. This function establishes a quantitative relationship with system parameters, enabling the discussion of the amplitude frequency response performance under different Rabi frequencies presented in this study. In addition, the power expression of the signal at the receiver describes the amplitude frequency response function applicable to general modulated signals. This formula intuitively illustrates the receiver bandwidth performance. The theoretical calculations and experimental measurement results are presented at the end of the paper. The paper also points out that the noise in the density matrix element $\rho_{ba}$ arises from the random distribution of population across energy levels. This noise, determined by the parameters of the receiver, cannot be eliminated. It is the intrinsic noise of the receiver and determines the ultimate sensitivity limit of the Rydberg atom receiver. This study overcomes the limitations of static solutions currently used to describe the four-level heterodyne receiver and establishes a more comprehensive theoretical framework.

Although the theoretical modeling in this paper adopts a detuning-free model, in practice, this model is applicable when the resonance frequencies of the probe and coupling lights maintain a specific locking relationship, given by \(\Delta_c = -\left(\lambda_p / \lambda_c\right)\Delta_p\). Under this condition, the resulting electromagnetically induced transparency effect is identical to that in the completely resonance (zero-detuning) case, which we refer to as equivalent resonance. The Rydberg receiver operating under equivalent resonance represents the most common receiver configuration, and our proposed method remains fully applicable in this scenario. Experimental measurements demonstrate that the response model of the atomic receiver established herein remains valid under this equivalent resonance condition.If the laser detunings do not satisfy the locking relationship, the method presented in this paper will not be applicable for analyzing the response of a Rydberg atomic receiver with arbitrary detunings. This scenario represents a topic for future research.

In our final system simulation, we assumed $\gamma_{{dep\_b}} = 0 $ and set $ \gamma_{{dep\_c}} = \gamma_{{dep\_d}}$, treating them as constants. However, this approach somewhat deviates from the actual physical scenario. The average number of collisions between atoms is related to the atomic concentration. The population concentration at each energy level is closely correlated with the receiver's parameters, meaning the dephasing decay rates for each level are also intimately connected to these receiver parameters. Consequently, high-precision modeling and estimation of the dephasing decay for each energy level represent an important area for future research.

In the case of direct detection of the transmitted light, due to the automatic down-conversion characteristics of the four-level heterodyne receiver, when the carrier frequency of the signal to be measured matches the local oscillator frequency, spectral aliasing occurs. Therefore, a proper frequency detuning between the two is required, and the signal bandwidth must not exceed the one-sideband of the down-converted center frequency. This limits the utilization of bandwidth resources. According to the dynamic solution theory, signal recovery can be achieved by utilizing both the real and imaginary parts of the susceptibility, enabling full utilization of the receiver's bandwidth. By using optical heterodyne detection technology, a reference light with a fixed frequency offset from the transmitted light is introduced. The two beams interfere at the detector to generate an intermediate-frequency signal, which can be demodulated orthogonally to simultaneously extract both the real and imaginary parts of the susceptibility, thereby achieving complete signal recovery. This approach effectively addresses the bandwidth limitation issue in direct detection.

\section*{Acknowledgments}
The authors acknowledge the financial support from the National Key Laboratory of Radar Signal Processing under Grant No. KGJ202101.

The authors extend their sincere gratitude to the reviewers for their invaluable contributions and diligent work, which have substantially improved the quality of this article.

\appendices
\section{Solving the Density Matrix Elements of a Four-Level System without the Field to be Measured}
\renewcommand{\theequation}{A\arabic{equation}}
The Hamiltonian matrix of the four-level system in the interaction picture is given by:
\setcounter{equation}{0}
\begin{equation}
H=\frac{\hbar}{2}
\begin{pmatrix}
0 & \Omega_pe^{i\Delta_pt} & 0 & 0 \\
\Omega_pe^{-i\Delta_pt} & 0 & \Omega_ce^{i\Delta_ct} & 0 \\
0 & \Omega_ce^{-i\Delta_ct} & 0 & \Omega_me^{i\Delta_mt} \\
0 & 0 & \Omega_me^{-i\Delta_mt} & 0
\end{pmatrix}.
\label{eqA1}
\end{equation}
Here, $\Delta_p$, $\Delta_c$ and $\Delta_m$ denote the detunings of the probe laser, coupling laser, and microwave field from their respective atomic transition frequencies. In the four-level heterodyne Rydberg atom receiver system, the probe, coupling, and local oscillator microwave fields are all set to resonate with their corresponding atomic transitions. By substituting Eq. (\ref{eqA1}) into the four-level density matrix equation 

\begin{equation}
\frac{d}{dt}\rho=  - \frac{i}{\hbar }\left( {H\rho  - \rho H} \right) - \frac{1}{2}\left( {\Gamma \rho  + \rho \Gamma } \right) + \Lambda.
\label{eqA2}
\end{equation}

A set of 16 coupled equations for the density matrix elements is obtained. To facilitate the solution, the density matrix elements are transformed as follows:
\begin{equation}
\begin{aligned}
 &{\rho}_{ab}^0=\rho_{ab}e^{-i\Delta_pt}, \rho_{ac}^0=\rho_{ac}e^{-i\left(\Delta_p+\Delta_c\right)t}, \\
 &\rho_{ad}^0=\rho_{ad}e^{-i\left(\Delta_p+\Delta_c+\Delta_m\right)t}, \rho_{ba}^0=\rho_{ba}e^{i\Delta_pt}, \\
 & \rho_{bc}^0=\rho_{bc}e^{-i\Delta_ct}, \rho_{bd}^0=\rho_{bd}e^{-i\left(\Delta_c+\Delta_m\right)t}, \\
 &{\rho}_{ca}^0=\rho_{ca}e^{i\left(\Delta_p+\Delta_c\right)t},\rho_{cb}^0=\rho_{cb}e^{i\Delta_ct}, \\
 &\rho_{cd}^0=\rho_{cd}e^{-i\Delta_mt}, {\rho}_{da}^0=\rho_{da}e^{i\left(\Delta_p+\Delta_c+\Delta_m\right)t}, \\ 
 &\rho_{db}^0=\rho_{db}e^{i\left(\Delta_c+\Delta_m\right)t}, \rho_{dc}^0=\rho_{dc}e^{i\Delta_mt},\rho_{ii}^0=\rho_{ii}.
\end{aligned}
\label{eqA3}
\end{equation}
The last equation in the system is replaced by $\rho_{aa}^0+\rho_{bb}^0+\rho_{cc}^0+\rho_{dd}^0=1$. Under the steady-state condition $d\rho_{ij}/dt=0$, the system reduces to a set of linear equations, which can be expressed in matrix form as: 
\begin{subequations}
\begin{equation}
    \mathbf{Ax}=\mathbf{y}.
    \label{eqA4}
\end{equation}
\begin{equation}
    \mathbf{x}={\begin{pmatrix}
   \rho _{aa}^{0} & \rho _{ab}^{0} & ... & \rho _{dc}^{0} & \rho _{dd}^{0}  \\
\end{pmatrix}}^{T},
\end{equation}
\begin{equation}
    \mathbf{y}={\begin{pmatrix}
   0 & 0 & ... & 0 & 1  \\
\end{pmatrix}}^{T}.
\end{equation}
\begin{equation}
\begin{aligned}
& A_{1,2}=-{{\Omega }_{p}}, A_{1,5}={{\Omega }_{p}}, A_{1,6}=2i{{\gamma }_{b}^{\prime}}, A_{1,11}=2i{\left(\gamma_t+\gamma_{dep\_c}\right)},\\
&A_{1,16}=2i{{\gamma }_{d}^{\prime}}; A_{2,1}=-{{\Omega }_{p}}, A_{2,2}=-i{{\gamma }_{ab}^{\prime}}+2{{\Delta }_{p}}, A_{2,3}=-{{\Omega }_{c}},\\
&A_{2,6}={{\Omega }_{p}};  A_{3,2}=-{{\Omega }_{c}}, A_{3,3}=-i{{\gamma }_{ac}^{\prime}}+2\left( {{\Delta }_{p}}+{{\Delta }_{c}} \right), \\ 
& A_{3,4}=-{{\Omega }_{m}}, A_{3,7}={{\Omega }_{p}}; A_{4,3}=-{{\Omega }_{m}}, \\
& A_{4,4}=-i{{\gamma }_{ad}^{\prime}}+2\left( {{\Delta }_{p}}+{{\Delta }_{c}}+{{\Delta }_{m}} \right), A_{4,8}={{\Omega }_{p}}; \\
& A_{5,1}={{\Omega }_{p}}, A_{5,5}=-i{{\gamma }_{ba}^{\prime}}-2{{\Delta }_{p}}, A_{5,6}=-{{\Omega }_{p}}, A_{5,9}={{\Omega }_{c}}; \\ 
& A_{6,2}={{\Omega }_{p}}, A_{6,5}=-{{\Omega }_{p}}, A_{6,6}=-2i{{\gamma }_{b}^{\prime}}, A_{6,7}=-{{\Omega }_{c}}, \\
& A_{6,10}={{\Omega }_{c}}, A_{6,11}=2i{{\gamma }_{c}}; A_{7,3}={{\Omega }_{p}}, A_{7,6}=-{{\Omega }_{c}}, \\ 
& A_{7,7}=-i{{\gamma }_{bc}^{\prime}}+2{{\Delta }_{c}}, A_{7,8}=-{{\Omega }_{m}}, A_{7,11}={{\Omega }_{c}}; \\ 
& A_{8,4}={{\Omega }_{p}}, A_{8,7}=-{{\Omega }_{m}}, A_{8,8}=-i{{\gamma }_{bd}^{\prime}}+2\left( {{\Delta }_{c}}+{{\Delta }_{m}} \right), \\
& A_{8,12}={{\Omega }_{c}}; A_{9,5}={{\Omega }_{c}}, A_{9,9}=-i{{\gamma }_{ca}^{\prime}}-2\left( {{\Delta }_{p}}+{{\Delta }_{c}} \right), \\ 
& A_{9,10}=-{{\Omega }_{p}}, A_{9,13}={{\Omega }_{m}}; A_{10,6}={{\Omega }_{c}}, A_{10,9}=-{{\Omega }_{p}}, \\
& A_{10,10}=-i{{\gamma }_{cb}^{\prime}}-2{{\Delta }_{c}}, A_{10,11}=-{{\Omega }_{c}}, A_{10,14}={{\Omega }_{m}}; \\ 
& A_{11,7}={{\Omega }_{c}}, A_{11,10}=-{{\Omega }_{c}}, A_{11,11}=-2i{{\gamma }_{c}^{\prime}}, \\
& A_{11,12}=-{{\Omega }_{m}}, A_{11,15}={{\Omega }_{m}}; A_{12,8}={{\Omega }_{c}}, A_{12,11}=-{{\Omega }_{m}}, \\
& A_{12,12}=-i{{\gamma }_{cd}^{\prime}}+2{{\Delta }_{m}}, A_{12,16}={{\Omega }_{m}}; A_{13,9}={{\Omega }_{m}}, \\ 
& A_{13,13}=-i{{\gamma }_{da}^{\prime}}-2\left( {{\Delta }_{p}}+{{\Delta }_{c}}+{{\Delta }_{m}} \right), A_{13,14}=-{{\Omega }_{p}};  \\
& A_{14,10}={{\Omega }_{m}}, A_{14,13}=-{{\Omega }_{p}}, \\
& A_{14,14}=-i{{\gamma }_{db}^{\prime}}-2\left( {{\Delta }_{c}}+{{\Delta }_{m}} \right), A_{14,15}=-{{\Omega }_{c}}; \\ 
& A_{15,11}={{\Omega }_{m}}, A_{15,14}=-{{\Omega }_{c}}, A_{15,15}=-i{{\gamma }_{dc}^{\prime}}-2{{\Delta }_{m}}, \\
& A_{15,16}=-{{\Omega }_{m}}; A_{16,1}=1, A_{16,6}=1, A_{16,11}=1, \\
& A_{16,16}=1. \\ 
\end{aligned}
\end{equation}
\end{subequations}
where $\gamma _{ij}^{\prime} = \gamma _i^{\prime} + \gamma_j^{\prime}$, $\gamma_a^{\prime}=\gamma_t$, $\gamma_b^{\prime}=\gamma_b+\gamma_t+\gamma_{dep\_b}$, $\gamma_c^{\prime}=\gamma_c+\gamma_t+\gamma_{dep\_c}$,
$\gamma_d^{\prime}=\gamma_d+\gamma_t+\gamma_{dep\_d}$.

In summary, the density matrix elements can be obtained by the inverse operation $\mathbf{x}=\mathbf{A}^{-1}\mathbf{y}$. It should be noted that the density matrix elements vary with atomic velocity due to Doppler shifts, which have been taken into account in the calculations presented in this work.

\section{Determination of the Phase Factor}
\setcounter{equation}{0}
\renewcommand{\theequation}{B\arabic{equation}}

Here, we take a single equation as an example to illustrate how the phase factor is determined. For the equation in (\ref{eq11}):
\begin{equation}
\begin{aligned}
&2i\left(\text{\small $\frac{\partial}{\partial t}$}+\mathbf{v}\cdot\nabla\right)\rho_{ba} = \Omega_{p}e^{ik_{p}z}\left(\rho_{aa}^{0}-\rho_{bb}^{0}\right)  \\
&\quad +\Omega_{c}e^{ik_{c}z}\rho_{ca}-i\gamma_{b}\rho_{ba}
\end{aligned},
\label{eqB1}
\end{equation}
when the microwave field to be measured is 0, we have ${{\rho }_{ca}}=\rho _{ca}^{\widetilde{0}}+e\left( t \right)=\rho _{ca}^{\widetilde{0}}$, ${{\rho }_{ba}}=\rho _{ba}^{\widetilde{0}}+e\left( t \right)=\rho _{ba}^{\widetilde{0}}$. Under this condition, the above equation can be rewritten as:
\begin{equation}
\begin{aligned}
&2i\left(\text{\small $\frac{\partial}{\partial t}$}+\mathbf{v}\cdot\nabla\right)\rho_{ba} = \Omega_{p}e^{ik_{p}z}\left(\rho_{aa}^{0}-\rho_{bb}^{0}\right)  \\
&\quad +\Omega_{c}e^{ik_{c}z}\rho_{ca}^{\widetilde{0}}-i\gamma_{b}\rho_{ba}^{\widetilde{0}}
\end{aligned}.
\label{eqB2}
\end{equation}
The term in the linear system (\ref{eqA4}) corresponding to Eq. (\ref{eqB2}) is given by:
\begin{equation}
{{\Omega }_{p}}\left( \rho _{aa}^{0}-\rho _{bb}^{0} \right)+{{\Omega }_{c}}\rho _{ca}^{0}-\left( i{{\gamma }_{b}}+2{{\Delta }_{p}} \right)\rho _{ba}^{0}=0.
\label{eqB3}
\end{equation}
By setting ${{\Delta }_{p}}=0$, it can be readily obtained from Eq. (\ref{eqB2}) that $\rho _{ba}^{\widetilde{0}}=\rho _{ba}^{0}{{e}^{i{{k}_{p}}z}}$ and $\rho _{ca}^{\widetilde{0}}=\rho _{ca}^{0}{{e}^{-i\left( {{k}_{c}}-{{k}_{p}} \right)z}}$.

Similarly, for the following equations:
\begin{equation}
{
\begin{array}{l}
2i\left( {\frac{\partial }{{\partial t}} + {\bf{v}} \cdot \nabla } \right){\rho _{ca}} = {\Omega _c}{e^{ - i{k_c}z}}{\rho _{ba}} + {\Omega _L}{\rho _{da}} - {\Omega _p}{e^{i{k_p}z}}{\rho _{cb}}\\
\begin{array}{*{20}{c}}
{\begin{array}{*{20}{c}}
{}&{}&{}
\end{array}}&{}&{}&{}
\end{array} + \frac{{\rho _{da}^{\widetilde 0}}}{{2\pi }}\int {{\Omega _s}\left( \omega  \right){e^{i\omega t}}d\omega }  - i{\gamma _c}{\rho _{ca}},
\end{array}
}
\end{equation}
\begin{equation}
\begin{array}{l}
2i\left( {\frac{\partial }{{\partial t}} + {\bf{v}} \cdot \nabla } \right){\rho _{cb}} = {\Omega _c}{e^{ - i{k_c}z}}\left( {\rho _{bb}^0 - \rho _{cc}^0} \right) + {\Omega _L}{\rho _{db}}\\
\begin{array}{*{20}{c}}
{\begin{array}{*{20}{c}}
{}&{}&{}
\end{array}}&{}&{}&{}
\end{array}{\rm{ + }}\frac{{\rho _{db}^{\widetilde 0}}}{{2\pi }}\int {{\Omega _s}\left( \omega  \right){e^{i\omega t}}d\omega }  - {\Omega _p}{e^{ - i{k_p}z}}{\rho _{ca}} - i{\gamma _{cb}}{\rho _{cb}}.
\end{array}
\end{equation}

when the microwave field to be measured is 0, we have ${{\rho }_{ba}}=\rho _{ba}^{\widetilde{0}}+e\left( t \right)=\rho _{ba}^{\widetilde{0}}$, ${{\rho }_{ca}}=\rho _{ca}^{\widetilde{0}}+e\left( t \right)=\rho _{ca}^{\widetilde{0}}$, ${{\rho }_{da}}=\rho _{da}^{\widetilde{0}}+e\left( t \right)=\rho _{da}^{\widetilde{0}}$, ${{\rho }_{cb}}=\rho _{cb}^{\widetilde{0}}+e\left( t \right)=\rho _{cb}^{\widetilde{0}}$, ${{\rho }_{db}}=\rho _{db}^{\widetilde{0}}+e\left( t \right)=\rho _{db}^{\widetilde{0}}$ and ${\Omega _s}\left( \omega  \right) = 0$. Under this condition, the above equation can be rewritten as:
\begin{equation}
\begin{array}{l}
2i\left( {\frac{\partial }{{\partial t}} + {\bf{v}} \cdot \nabla } \right){\rho _{ca}} = {\Omega _c}{e^{ - i{k_c}z}}\rho _{ba}^{\widetilde 0} + {\Omega _L}\rho _{da}^{\widetilde 0}\\
\begin{array}{*{20}{c}}
{\begin{array}{*{20}{c}}
{}&{}&{}&{}
\end{array}}&{}&{}&{}
\end{array} - {\Omega _p}{e^{i{k_p}z}}\rho _{cb}^{\widetilde 0} - i{\gamma _c}\rho _{ca}^{\widetilde 0},
\end{array}
\label{eqB6}
\end{equation}
\begin{equation}
\begin{array}{l}
2i\left( {\frac{\partial }{{\partial t}} + {\bf{v}} \cdot \nabla } \right){\rho _{cb}} = {\Omega _c}{e^{ - i{k_c}z}}\left( {\rho _{bb}^0 - \rho _{cc}^0} \right) + {\Omega _L}\rho _{db}^{\widetilde 0}\\
\begin{array}{*{20}{c}}
{\begin{array}{*{20}{c}}
{}&{}&{}&{}
\end{array}}&{}&{}&{}
\end{array} - {\Omega _p}{e^{ - i{k_p}z}}\rho _{ca}^{\widetilde 0} - i{\gamma _{cb}}\rho _{cb}^{\widetilde 0}.
\end{array}
\label{eqB7}
\end{equation}

The term in the Eq. (\ref{eqA4}) corresponding to Eq. (\ref{eqB6}) and Eq. (\ref{eqB7}) are given by:
\begin{equation}
{\Omega _c}\rho _{ba}^0 + {\Omega _L}\rho _{da}^0 - \rho _{cb}^0{\Omega _p} - \left( {i{\gamma _c} + 2\left( {{\Delta _p} + {\Delta _c}} \right)} \right)\rho _{ca}^0{\rm{ = }}0,
\end{equation}
\begin{equation}
{\Omega _c}\left( {\rho _{bb}^0 - \rho _{cc}^0} \right) + {\Omega _L}\rho _{db}^0 - \rho _{ca}^0{\Omega _p} - \left( {i{\gamma _{bc}} + 2{\Delta _c}} \right)\rho _{cb}^0{\rm{ = }}0.
\end{equation}

By setting ${{\Delta }_{p}}=0$ and ${{\Delta }_{c}}=0$, it can be readily obtained that $\rho _{da}^{\widetilde{0}}=\rho _{da}^{0}{{e}^{-i\left(k_c-k_p\right)z}}$, $\rho _{cb}^{\widetilde{0}}=\rho _{cb}^{0}{{e}^{-i{{k}_{c}} z}}$ and $\rho _{db}^{\widetilde{0}}=\rho _{db}^{0}{{e}^{-i{{k}_{c}} z}}$.
\bibliographystyle{IEEEtran}
\bibliography{references}

\end{document}